\documentclass{article}
\usepackage{iclr2026_conference,times}

\usepackage{amsmath,amsfonts,bm}

\def\eqref#1{equation~\ref{#1}}

\def\1{\bm{1}}

\DeclareMathAlphabet{\mathsfit}{\encodingdefault}{\sfdefault}{m}{sl}
\SetMathAlphabet{\mathsfit}{bold}{\encodingdefault}{\sfdefault}{bx}{n}

\usepackage{hyperref}
\usepackage{url}
\usepackage{graphicx}
\usepackage{wrapfig}
\usepackage{booktabs}
\usepackage{multirow}
\usepackage[table]{xcolor}
\usepackage{colortbl}
\usepackage{fancyhdr}
\usepackage{enumitem}
\usepackage{wrapfig}
\usepackage{caption}
\usepackage{lipsum}
\usepackage{subcaption} 
\usepackage{float}
\usepackage{array}
\usepackage{tabularx}
\usepackage{makecell}
\title{RDDM: Practicing RAW Domain Diffusion Model for Real-world Image Restoration}
\author{
Yan Chen$^{1}$$^{\ast}$, Yi Wen$^{1}$\thanks{Equal Contribution.}  , Wei Li$^{1}$\thanks{Project Lead.}  , Junchao Liu$^{1}$, Yong Guo$^{2}$, Jie Hu$^{1}$, Xinghao Chen$^{1}$\\
$^{1}$ Huawei Noah's Ark Lab \\
$^{2}$ Max Planck Institute for Informatics \\
\texttt{\{chenyan176, wenyi14, wei.lee\}@huawei.com}
}

\iclrfinalcopy
\begin{document}

\maketitle
\begin{figure*}[h]
    \centering
    \includegraphics[width=\textwidth]{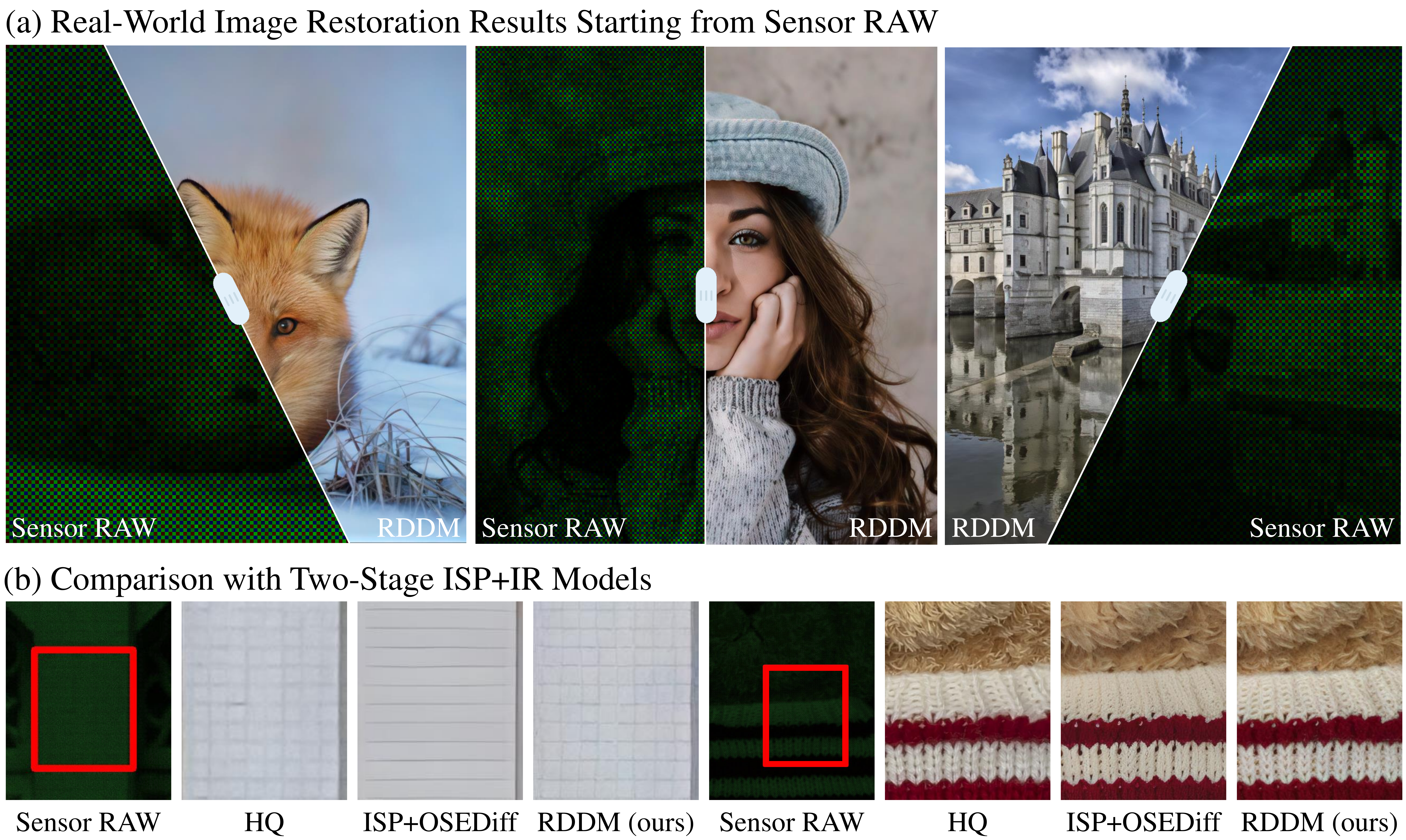}
    \caption{RDDM, restoring directly from the sensor RAW data, demonstrates remarkable results shown in (a), capitalizing on the unprocessed and detail-rich signal. Compared with the two-stage baseline in (b), RDDM delivers markedly higher fidelity and perceptual quality.}
    \label{fig:Overview}
\end{figure*}

\begin{abstract}
We present the RAW domain diffusion model (RDDM), an end-to-end diffusion model that restores photo-realistic images directly from the sensor RAW data. While recent sRGB-domain diffusion methods achieve impressive results, they are caught in a dilemma between high fidelity and image generation. As these models process lossy sRGB inputs and neglect the accessibility of the sensor RAW images in many scenarios, e.g., in image and video capturing in edge devices, resulting in sub-optimal performance. RDDM obviates this limitation by directly restoring images in the RAW domain, replacing the conventional two-stage image signal processing (ISP)$\rightarrow$Image Restoration (IR) pipeline. However, a simple adaptation of pre-trained diffusion models to the RAW domain confronts many challenges. To this end, we propose: (1) a RAW-domain VAE (RVAE), encoding sensor RAW and decoding it into an enhanced linear domain image, to solve the out-of-distribution (OOD) issues between the different domain distributions; (2) a configurable multi-bayer (CMB) LoRA module, adapting diverse RAW Bayer patterns such as RGGB, BGGR, etc. To compensate for the deficiency in the dataset, we develop a scalable data synthesis pipeline synthesizing RAW LQ-HQ pairs from existing sRGB datasets for large-scale training. Extensive experiments demonstrate RDDM's superiority over state-of-the-art sRGB diffusion methods, yielding higher fidelity results with fewer artifacts. Codes will be publicly available at  \href{https://github.com/YanCHEN-fr/RDDM}{github.com/YanCHEN-fr/RDDM}.

\end{abstract}

\section{Introduction}

Real-world Image Restoration (Real-IR) aims to restore high-quality (HQ) images from low-quality (LQ) images containing complex degradations \cite{fan2020neural,zhang2022accurate, jinjin2020pipal, zhang2019residual, zhang2018residual, zhang2017learning}. Real-IR has transitioned from GAN-based adversarial training \cite{ledig2017photo,wang2018esrgan,xie2023desra} to leveraging the powerful generative priors of text-to-image (T2I) diffusion models \cite{zhang2023adding,lin2024diffbir,wang2024exploiting,wu2024seesr,yu2024scaling}. Despite their success in the sRGB domain, these methods encounter a fundamental bottleneck: the Image Signal Processor (ISP) irreversibly discards critical details and compresses the dynamic range from high-bit (12-14 bits) RAW images into 8-bit sRGB images. Restoring HQ images from such degraded sRGB inputs inevitably leads to sub-optimal results \cite{nguyen2016raw, xing2021invertible}.

\begin{figure*}[t]
    \centering
    \begin{minipage}[t]{0.42\linewidth}
    \captionsetup{type=figure}
      \centering
      \includegraphics[width=\linewidth]{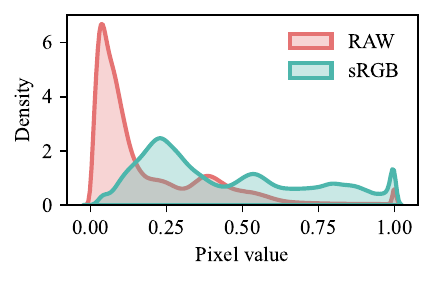}
      \caption{Distribution gap between RAW and sRGB images.}
      \label{fig:distribution shift}
    \end{minipage}
    \hfill
    \hspace{1em}
    \begin{minipage}[t]{0.53\linewidth}
    \captionsetup{type=figure}
      \centering
      \includegraphics[width=\linewidth]{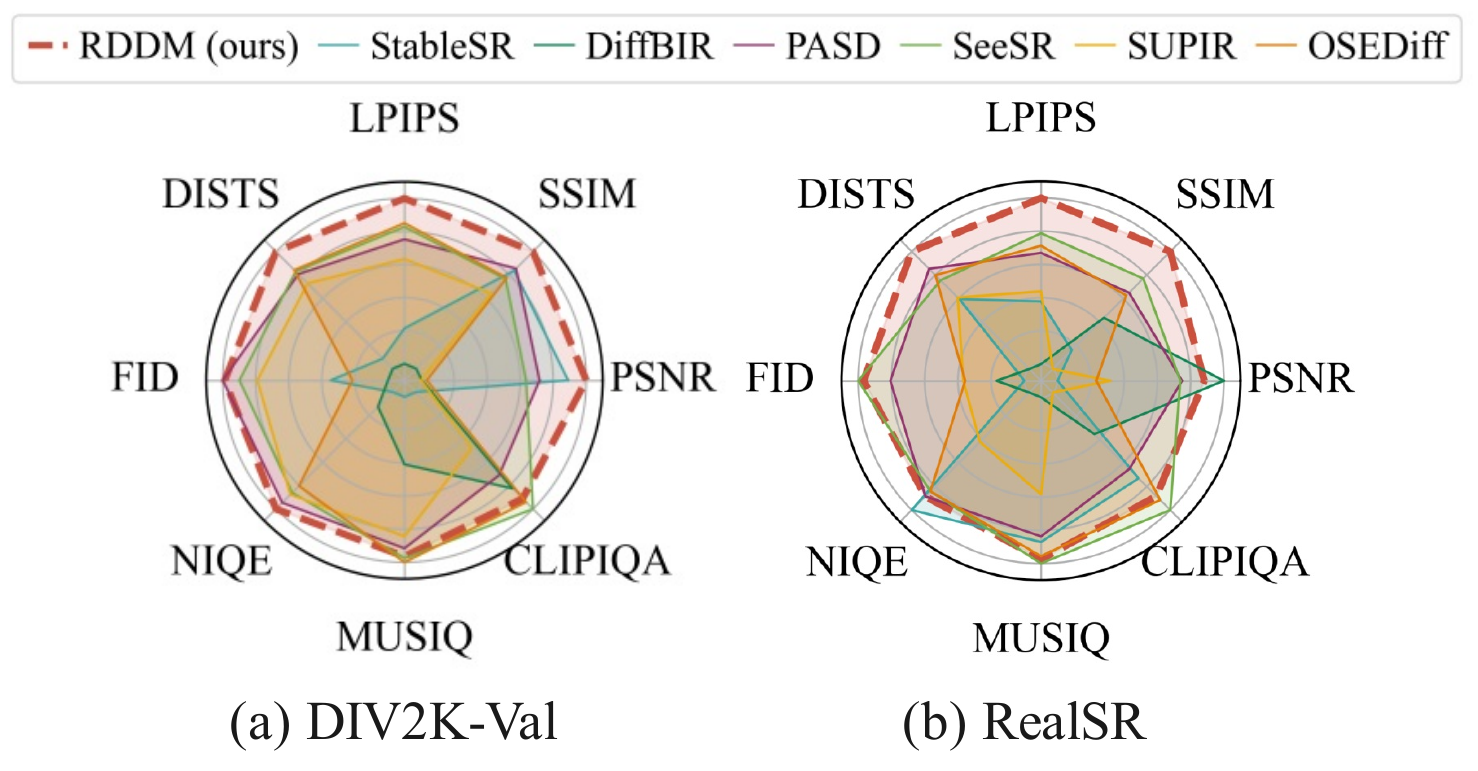}
      \caption{The performance comparison among SD-based methods on test datasets DIV2K-Val, and RealSR, respectively.}
      \label{fig:radar}
    \end{minipage}
\end{figure*}

\noindent To transcend the performance ceiling of Real-IR, shifting restoration to the sensor RAW domain, i.e. the front-end of ISP, holds substantial promises. RAW data preserves the original optical and physical signals, offering higher bit depth and richer information. Nevertheless, integrating T2I models into the RAW domain is confronted with two critical challenges: (1) \textbf{Distribution Disparity:} the significant gaps in the luminance and mosaic patterns between the sRGB and RAW images, as shown in Fig.\ref{fig:distribution shift}, make direct application of T2I models in the RAW domain ineffective. (2) \textbf{Generalization Constraints:} varying sensor Bayer patterns and noise characteristics hinder the ability of the model to generalize across different devices.

\noindent In this paper, we propose the RAW Domain Diffusion Model (RDDM), the first practical diffusion-based paradigm for RAW image restoration. To address the aforementioned challenges, we introduce three core innovations: 

\begin{itemize}
\item \textbf{RVAE:} we design a RAW domain Variational Autoencoder (RVAE) using a divide-and-conquer strategy, mapping noisy mosaicked RAW inputs into a latent space and decoding them into enhanced linear images to bridge the RAW-sRGB distribution gap.
\item \textbf{CMB-LoRA:} a sensor-aware adaptation module that strategically selects LoRA weights to handle heterogeneous Bayer patterns, significantly enhancing cross-sensor generalization.
\item \textbf{RAW Image Synthesis Pipeline:} to overcome data scarcity, we propose a synthesis method to generate high-fidelity RAW-linear pairs from large-scale sRGB datasets.
\end{itemize}

\noindent Extensive experiments demonstrate that RDDM achieves state-of-the-art fidelity and generative performance, as illustrated in Fig.~\ref{fig:radar}.

\section{Related Work}
\textbf{Real-world Image Restoration.} Since the advent of ESRGAN \cite{ledig2017photo}, Real-IR has evolved from GAN-based approaches \cite{ledig2017photo,wang2018esrgan,wang2021real,zhang2021designing,liang2022details,chen2022real}, which often suffer from training instability and unnatural artifacts, to diffusion-based models. By leveraging pre-trained Stable Diffusion (SD) priors, the latter achieve superior visual realism and texture generation through various condition injection strategies \cite{podell2023sdxl,rombach2022high,wang2024exploiting,wu2024seesr,yu2024scaling}. However, existing Real-IR methods perform predominantly in the sRGB domain, where critical sensor-level information is often discarded by the ISP. Furthermore, directly transferring sRGB-trained models to the RAW domain yields poor performance due to severe domain mismatch. This gap necessitates a specialized framework that effectively adapts the generative power of diffusion models to the unique characteristics of RAW data.

\noindent\textbf{Image Processing in RAW Domain.} Modern imaging sensors capture RAW data, which is converted to sRGB by a traditional ISP pipeline via sequential modules, including demosaicing (DM), denoising (DN), and tone mapping (TM). Subsequent post-processing (e.g., AWB, clipping) compresses dynamic range, leading to irreversible detail loss in the final sRGB output. One-stage methods \cite{brooks2019unprocessing} typically integrate DN and DM but often suffer from over-smoothing due to limited representation capacity. In contrast, integrating diffusion models into the RAW domain harnesses essential sensor signals and enhances generative capacity, providing a superior foundation for high-fidelity image restoration.

\section{Methodology}
\subsection{Problem Modeling} 
\label{sec:3.1}
In the sRGB domain, Real-IR model $G_{\theta}^{rgb}$, parameterized by $\theta$, aims to estimate HQ sRGB image $\hat{X}_{H}^{rgb} \in \mathbb{R}^{h \times w\times 3}$ given LQ sRGB image $X_{L}^{rgb} \in \mathbb{R}^{h \times w\times 3}$. In RAW domain, we train a neural network $G_{\theta}^{raw}$ to transform LQ sensor RAW $X_{L}^{raw} \in \mathbb{R}^{h \times w\times 1}$ to HQ linear domain image $\hat{X}_{H}^{lin} \in \mathbb{R}^{h \times w\times 3}$. The training task can be modeled as the following optimization problem: 

{\small
\begin{equation}\label{training_optimization_problem}
\theta^{*} = argmin_{\theta}\mathbb{E}_{X_{L}^{raw}, X_{H}^{lin} \sim S} \left[\mathcal{L}(G_{\theta}^{raw}(X_{L}^{raw}, \epsilon), X_{H}^{lin}) \right]
\end{equation}
}
where S is the dataset consisting of ($X_{L}^{raw}, X_{H}^{lin}$) pairs, and $\mathcal{L}$ is the loss function, respectively.

\begin{figure}[t]
  \centering
  \includegraphics[width=\linewidth]{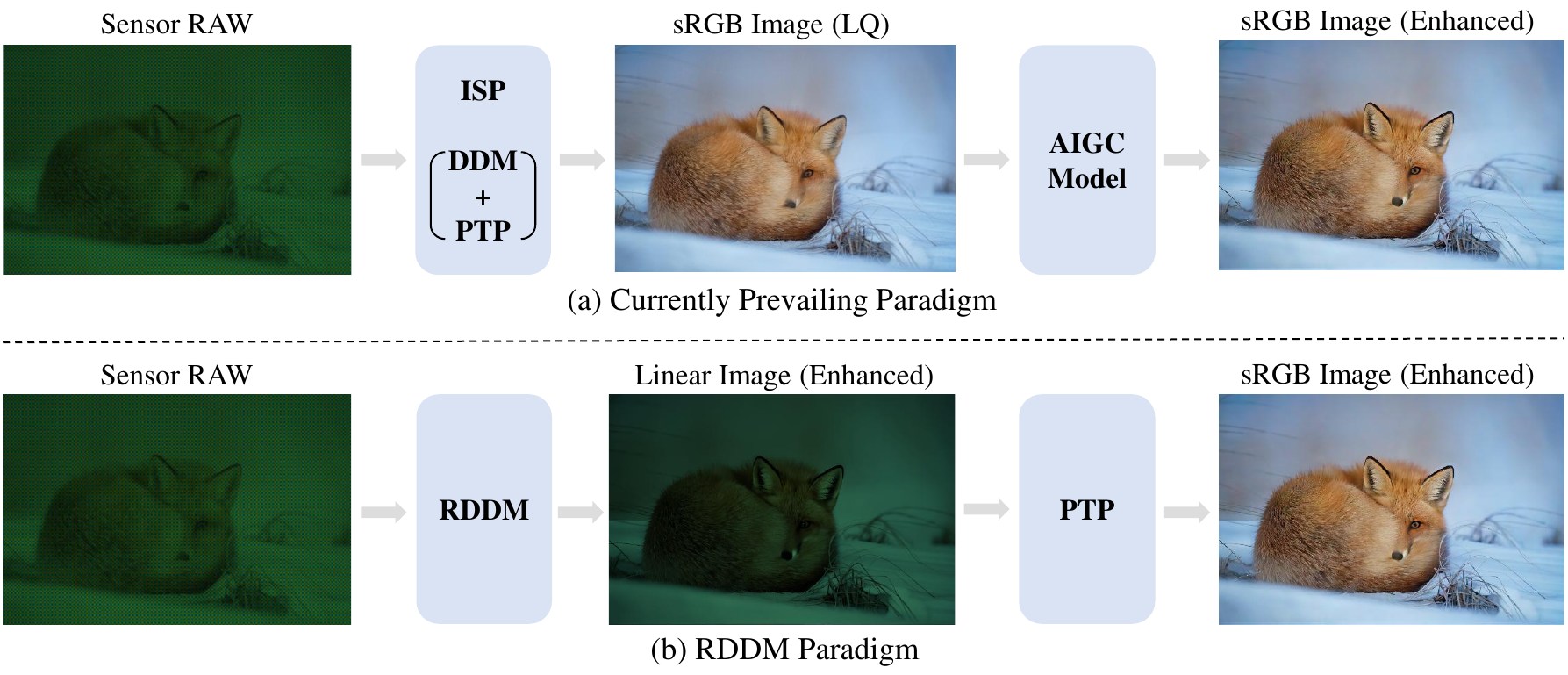}
  \caption{Comparison of Real-IR paradigms. (a) The prevailing paradigm: the restoration process is performed in the sRGB space after the Raw-to-sRGB mapping via an ISP. (b) RDDM paradigm: we perform restoration directly on the RAW data and subsequently map the restored results to the sRGB space using a PTP module.}
  \label{fig:probem_modeling}
\end{figure}

\noindent Existing AIGC-based image restoration paradigms typically follow the pipeline illustrated in Fig.~\ref{fig:probem_modeling} (a). The initial sensor RAW is processed through conventional image signal processing, which integrates a denoising, demosaicing module (DDM), and post-tone mapping modules to generate an sRGB image. The AIGC model is subsequently applied to this sRGB output to perform high-quality enhancement. Formally, this paradigm can be expressed as:

\begin{equation} \label{denoisin demosaicing}
    \hat{X}_H^{rgb} = G_{\theta}^{rgb}(ISP(X_{L}^{raw}))
\end{equation}

\noindent In contrast, our proposed RDDM approach, depicted in Fig.~\ref{fig:probem_modeling} (b), bypasses the traditional denoising and demosaicing modules and shifts the AIGC-driven enhancement process into the RAW domain. Following the enhancement, the output of the AIGC model is transformed into the sRGB domain via a post tone mapping (PTP) module. Formally, the RDDM paradigm can be defined as:

\begin{equation} \label{denoisin demosaicing}
    \hat{X}_H^{rgb} = PTP(G_{\theta}^{raw}(X_{L}^{raw}))
\end{equation}

\noindent By operating directly on RAW data, RDDM mitigates the irreversible information loss inherent in standard ISP transformations. Since on-device ISP pipelines are typically proprietary "black boxes," a comprehensive overview of ISP components and their corresponding mathematical derivations is provided in Appendix ~\ref{sec:A1}.

\subsection{RAW Domain Diffusion Model} \label{section:RDDM}

\begin{figure*}[t]
  \centering
  \includegraphics[width=\textwidth]{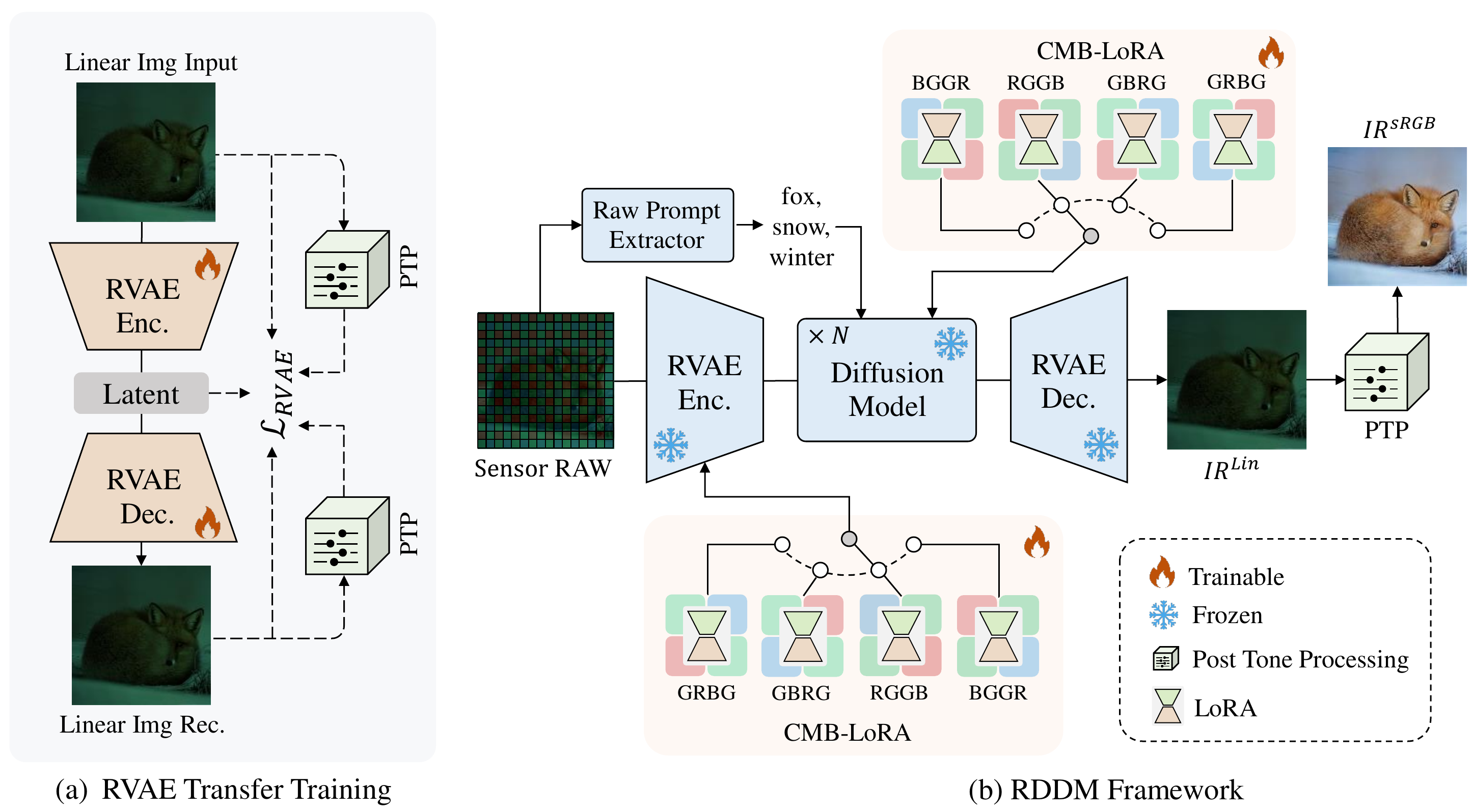}
  \caption{ (a) Illustration of RVAE training strategy. (b) RDDM framework. Specifically, it synchronizes the optimization of the CMB-LoRA (within the RVAE encoder) and the pre-trained diffusion network, supervised by RAW-linear image pairs.}
  \label{fig:achitecture}
\end{figure*}

\textbf{Framework Overview}. Our framework $G_{\theta}^{raw}$ integrates an RVAE and a diffusion network $\epsilon_{\theta}$ for high-fidelity restoration. Specifically, the encoder $E_{\theta}^{lin}$ maps mosaicked RAW inputs into a latent space, where $\epsilon_{\theta}$ performs iterative refinement to recover fine-grained details. The decoder $D_{\theta}^{lin}$ then reconstructs these features into the linear domain, followed by an sRGB transformation via the PTP module. To handle diverse Bayer configurations, we incorporate trainable LoRA layers \cite{hu2022lora} into the pre-trained $E_{\theta}^{lin}$ and $\epsilon_{\theta}$, facilitating adaptive adaptation through our CMB-LoRA strategy. Furthermore, to provide semantic guidance, we employ a DAPE \cite{zheng2024dape} to derive textual prompts from sRGB images rendered by the ISP, thereby activating generative priors. An overview of the framework is depicted in Fig.~\ref{fig:achitecture}.

\noindent\textbf{RAW Domain VAE.} Owing to the vast domain shift, existing sRGB domain VAEs fail to reconstruct effectively, resulting in diffusion model-generated images characterized by low visual quality and significant color shifts. Therefore, we devise a RAW domain VAE that encodes RAW images and subsequently decodes the latent representation into a linear domain image. The training of the RVAE is conducted in two distinct stages. To ensure the decoder is capable of reconstructing images in the linear domain, the first stage involves training on a reconstruction task utilizing linear domain images, following the strategy illustrated in Fig.~\ref{fig:achitecture} (a). Given a linear domain input $X_{H}^{lin}$, the encoder extracts the latent representation $z$, which is then reconstructed by the decoder as $\hat{X}_{H}^{lin} = D_{\theta}^{lin}(E_{\theta}^{lin}(X_{H}^{lin}))$. Furthermore, we devise a differentiable PTP module that simultaneously supervises training in both the sRGB and RAW domains. Similar to LDM \cite{rombach2022high}, we use $L_1$
loss, LPIPS loss, and GAN loss to train the VAE encoder and decoder to generate realistic details of a linear image:

\begin{equation} \label{}
\begin{split}
    \small
    \mathcal{L}_{RVAE} = \mathcal{L}_{rec} (\hat{X}_{H}^{lin}, X_{H}^{lin}) + \lambda_{G} \mathcal{L}_{GAN}(\hat{X}_{H}^{lin}, X_{H}^{lin})
\end{split}
\end{equation}
where $\mathcal{L}_{rec} = \mathcal{L}_{1} + \mathcal{L}_{LPIPS}$ and $\mathcal{L}_1$ is calculated in both the RAW and sRGB domains. $\lambda_{G} = \frac{\nabla[L_{rec}]}{\nabla[L_{GAN}]+10^{-4}}$ and $\nabla[\cdot]$ represents the gradient of the last layer in the decoder. Following \cite{rombach2022high}, to facilitate diffusion model optimization, we normalize the latent distribution to a standard Gaussian. The scaling factor $\sigma$ is computed over the training samples:

\begin{align} \label{equation:scaling_factor}
    \hat{\mu} = \frac{1}{bchw} \sum_{b,c,h,w} z^{b,c,h,w}
\end{align}

\begin{align} \label{equation:scaling_factor}
    \sigma^{2} = \frac{1}{bchw} \sum_{b,c,h,w} (z^{b,c,h,w} - \hat{\mu})^2
\end{align}
where $z^{b,c,h,w}$ denotes the latent space of the training samples encoded by $E_{\theta}^{lin}$. $\hat{\mu}$ and $\sigma^{2}$ present the mean and variance of the data distribution. The rescaled latent has unit standard deviation, i.e., $z \leftarrow \frac{z}{\sigma}$. To enable the RVAE to process RAW images, the second training stage involves freezing the encoder and decoder weights optimized in the first stage and incorporating LoRA into the encoder. Concurrently, LoRA modules are integrated into the diffusion model to facilitate the removal of mosaic artifacts and heavy noise while simultaneously enhancing fine-grained image details. The overall training strategy is illustrated in Fig.~\ref{fig:achitecture} (b). We use VSD loss, LPIPS loss, and MSE loss to train our model in the RAW domain and the sRGB domain:

\begin{equation}
\label{equation:l_data}
\begin{aligned} 
    \mathcal{L} &= \mathcal{L}_{VSD}(\hat{X}_{H}^{lin}, X_{H}^{lin}) + \lambda_{1} \mathcal{L}_{RAW}(\hat{X}_{H}^{lin}, X_{H}^{lin})  \\ 
    &+ \lambda_{2} \mathcal{L}_{RGB}(\mathcal{F}_{PTP}(\hat{X}_{H}^{lin}), \mathcal{F}_{PTP}(X_{H}^{lin}))
\end{aligned}
\end{equation}
where $\lambda_{1}$, $\lambda_{2}$ are weighting scalars. $\mathcal{L}_{RAW} = \mathcal{L}_{MSE}$. $\mathcal{L}_{RGB} = \mathcal{L}_{MSE} + \lambda_{3} \mathcal{L}_{LPIPS}$. We transform both the model predictions and the ground-truth into the sRGB domain via the proposed PTP module.

\noindent\textbf{CMB-LoRA.} 
To tackle the traditional cross-device generalization challenge posed by diverse Bayer patterns and noise characteristics, we propose Configurable Multi-Bayer (CMB) LoRA. Rather than costly full-parameter retraining, our approach incorporates pattern-specific LoRA branches into the RVAE and diffusion networks. Each branch is independently optimized for a specific Bayer configuration during training and adaptively selected during inference. This parameter-efficient strategy allows our model to handle diverse sensor characteristics without retraining from scratch, as detailed in Fig.~\ref{fig:achitecture} (b).

\begin{figure}[t]
  \centering
  \includegraphics[width=0.6\linewidth]{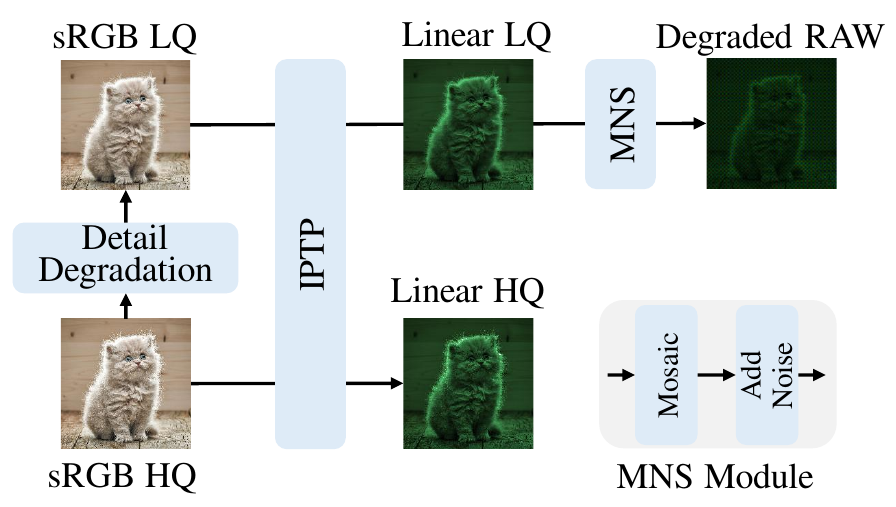}
  \caption{RAW data synthesis pipeline. IPTP transforms an sRGB image into its linear domain counterpart. MNS converts a linear domain image into a sensor RAW image.}
  \label{fig:isp_data_synthesis}
\end{figure}

\subsection{Data Synthesis} \label{section:image_generalization}
Despite the prevalence of sRGB datasets for Real-IR (e.g., LSDIR \cite{li2023lsdir}, DIV2K \cite{agustsson2017ntire}), a large-scale RAW domain dataset remains absent. To address this deficiency, we construct a large-scale synthetic RAW-domain dataset by applying a degradation pipeline to publicly available sRGB resources, thereby facilitating effective training in the RAW domain, as shown in Fig.~\ref{fig:isp_data_synthesis}. Following Real-ESRGAN \cite{wang2021real}, we first apply detail degradation to obtain $X_{L}^{rgb}$ from $X_{H}^{rgb}$. Crucially, we exclude sRGB-specific JPEG compression and random noise, as these do not align with native sensor characteristics. For stage one training, an inverse post tone processing (IPTP) module projects $X_{H}^{rgb}$ into the linear domain to obtain the RVAE target $X_{H}^{lin}$. In stage two, we synthesize RAW domain LQ inputs $X_{L}^{raw}$ by passing $X_{L}^{rgb}$ through the IPTP module and a mosaic noise synthesizer (MNS). Throughout training, $X_{H}^{lin}$ serves as the ground truth. The process is formulated as:

\begin{equation} \label{denoisin demosaicing}
    X_H^{lin} = IPTP(X_{H}^{rgb}), X_L^{raw} = MNS(IPTP(X_{H}^{lin}))
\end{equation}

where $X_{H}^{lin}$ provides the supervision signal for the reconstruction of $X_{L}^{raw}$. A comprehensive overview of data synthesis is provided in Appendix~\ref{sec:A1} and some visual examples of the synthetic data are found in Appendix ~\ref{sec:A7}.

\section{Experiments}

\begin{table*}[t]
    \caption{Quantitative comparison with different methods on both synthetic benchmarks. The best, second best and third results of each metric are highlighted by \colorbox{red!40}{red}, \colorbox{orange!40}{orange} and \colorbox{yellow!40}{yellow} cells respectively. $\downarrow$ presents the smaller the better, $\uparrow$ presents the bigger the better. Please note that we denote the number of sampling steps for each diffusion-based method using the format "method-steps".}

    \centering
    \setlength{\tabcolsep}{1.7mm}
    \scriptsize
    \begin{tabular}{c|c|cccccccc}
        \toprule
        \textbf{Dataset} & \textbf{Method} & \textbf{PSNR$\uparrow$} & \textbf{SSIM$\uparrow$} & \textbf{LPIPS$\downarrow$} & \textbf{DISTS$\downarrow$} & \textbf{FID$\downarrow$} & \textbf{NIQE$\downarrow$} & \textbf{MUSIQ $\uparrow$} & \textbf{CLIPIQA $\uparrow$}\\
        \midrule
        \multirow{11}*{DIV2K-Val} & JDnDmSR & {23.4565}  & \cellcolor{yellow!40}0.6192  & {0.5347}  & {0.2655}  & {45.3706}  & {7.0895}  & {32.1252}   & {0.1978} \\
        ~ & SwinIR & 22.7983  & \cellcolor{orange!40}{0.6294}  & {0.5345}  & {0.2780}  & {44.9270}  & {7.1012}  & {32.9053}  & {0.2520} \\
        ~ & MambaIRv2  &  \cellcolor{orange!40}{23.6377} & 0.6009  &  0.5882 & 0.2749  &  42.8625 & 7.2673  & 31.6576  & 0.2010 \\ \cline{2-10}
        ~ & ISP+StableSR-s200 & \cellcolor{yellow!40}{23.6034}  & 0.6133  & 0.4095  & 0.2092  & 35.6300 & {4.7840}  & {43.8325}  & {0.4284}  \\
        ~ & ISP+DiffBIR-s50 & {22.4903}  & {0.5284}  & {0.4519}  & {0.2176}  & {42.0167}  & 4.6040  & 52.9640  & 0.6503  \\
        ~ & ISP+PASD-s20 & 23.3860 & {0.6150}  &  0.3029 &  0.1385 &   \cellcolor{red!40}{23.5801} & \cellcolor{orange!40}{3.4392}  & 64.3181 & 0.6197  \\
        ~ & ISP+SeeSR-s50 & 23.2836  & {0.6059}  & \cellcolor{yellow!40}{0.2880}  & \cellcolor{yellow!40}{0.1363}  & \cellcolor{yellow!40}{25.4424}  & 3.5605  & \cellcolor{orange!40}{65.6650}  & \cellcolor{red!40}{0.6976}  \\
        ~ & ISP+SUPIR-s50 & {22.4837}  & {0.5935}  & 0.3265  & 0.1462  & 27.4418  & \cellcolor{yellow!40}{3.5376}  & 62.7078  & 0.5570  \\
        ~ & ISP+OSEDiff-s1 & {22.5277} & 0.6069 & \cellcolor{orange!40}{0.2836} & \cellcolor{orange!40}{0.1351} & 38.0461 & 3.6427 & \cellcolor{red!40}{66.2024} & \cellcolor{orange!40}{0.6818} \\
        ~ & Ours-s1 & \cellcolor{red!40}{23.7416} & \cellcolor{red!40}{0.6296} &  \cellcolor{red!40}{0.2540} & \cellcolor{red!40}{0.1197}  & \cellcolor{orange!40}{23.8028}  & \cellcolor{red!40}{3.3627}  & \cellcolor{yellow!40}{65.4202}    & \cellcolor{yellow!40}{0.6737}  \\
        \midrule
        \multirow{11}*{DRealSR} & JDnDmSR & 27.6972   & \cellcolor{orange!40}{0.7995}   & {0.3610}   & {0.2210}   & {31.1697}   & {7.9294}   & {30.4728}  & {0.2373}  \\
        ~ & SwinIR  & {27.0657}  & \cellcolor{red!40}{0.8161}   & {0.3714}   & {0.2305}   & {30.5639}   & {7.5234}   & {30.2268}    & {0.2972}  \\
        ~ & MambaIRv2  &  \cellcolor{red!40}{28.5563} & 0.7655  &  0.4409 & 0.2359  & 27.8182  &  8.1602 & 29.1201  & 0.2595 \\ \cline{2-10}
        ~ & ISP+StableSR-s200  & 27.1173  & 0.7613  & 0.3387  & 0.1978  & 25.8442  & \cellcolor{orange!40}{4.5959}  & 49.2604  & 0.5991  \\
        ~ & ISP+DiffBIR-s50  & 28.2670  & {0.7606}  & {0.4142}  & {0.2702}  & 25.9530  & {6.3725}  & {38.1396}  & 0.5284  \\
        ~ & ISP+PASD-s20 & \cellcolor{yellow!40}{28.3377} & {0.7845}  & \cellcolor{orange!40}{0.2870}  &  \cellcolor{orange!40}{0.1670} &  \cellcolor{red!40}{16.1714} &  4.6875 & 53.1539 & 0.5872  \\
        ~ & ISP+SeeSR-s50 & 27.6513 & 0.7765  & \cellcolor{yellow!40}{0.2972}  & 0.1816  & \cellcolor{yellow!40}19.2938  & \cellcolor{red!40}{4.2053}  & \cellcolor{yellow!40}{56.0800}  & \cellcolor{yellow!40}{0.6681}  \\
        ~ & ISP+SUPIR-s50  & {26.9559}  & {0.7359}  & 0.3262  & \cellcolor{yellow!40}{0.1799}  & 26.1866  & 5.0892  & 48.5114  & {0.4839}  \\
        ~ & ISP+OSEDiff-s1 & {25.1101}  & {0.7315}  & 0.3396  & 0.1900  & {32.4002}  & 4.7336  & \cellcolor{red!40}{57.3375}  & \cellcolor{red!40}{0.7376}  \\
        ~ & Ours-s1 &   \cellcolor{orange!40}{28.3495} &  \cellcolor{yellow!40}{0.7892} & \cellcolor{red!40}{0.2719}  & \cellcolor{red!40}{0.1649}  &  \cellcolor{orange!40}{17.4825} & \cellcolor{yellow!40}{4.6852}  & \cellcolor{orange!40}{57.0696}  &  \cellcolor{orange!40}{0.7035} \\
        \midrule
        \multirow{11}*{RealSR} & JDnDmSR & \cellcolor{orange!40}{25.6346}  & \cellcolor{red!40}{0.7532}  & {0.3649}  & {0.2119}  & {66.5709}  & {7.4103}  & {38.6661} & {0.2062} \\
        ~ & SwinIR & \cellcolor{yellow!40}{25.4564}  & \cellcolor{orange!40}{0.7477}  & {0.3818}  & {0.2283}  & {67.1467}  & {6.9218}  & {38.5181}   & {0.2637} \\
        ~ & MambaIRv2 & \cellcolor{red!40}{25.6781}  & 0.6976  &  0.4686 & 0.2399  & 64.8715  & 7.4959  & 36.2515  & 0.2100 \\ \cline{2-10}
        ~ & ISP+StableSR-s200 &  {23.3339} &  {0.6600} & 0.3505  & 0.1949  &  {60.9322} & \cellcolor{red!40}{3.9343}  &  64.1478 &  0.6393 \\
        ~ & ISP+DiffBIR-s50 &  {25.3643} & {0.6761} &  {0.4086} & {0.2478}  & 56.7401  & {5.6140}  & {49.4878}  & 0.5581 \\
        ~ & ISP+PASD-s20 & 24.8545  & 0.6886  & 0.3055  &  \cellcolor{orange!40}{0.1720} & \cellcolor{yellow!40}{40.8756}  & \cellcolor{yellow!40}{4.1290}  & 63.5759  & 0.6223  \\
        ~ & ISP+SeeSR-s50 & 24.8332  & {0.6957}  & \cellcolor{orange!40}{0.2872}  & 0.1807  &  \cellcolor{red!40}{36.0702} &  4.2017 &  \cellcolor{red!40}{66.3191}  & \cellcolor{red!40}{0.6977}  \\
        ~ & ISP+SUPIR-s50 & {23.9782}  & {0.6505}  &  0.3412 & 0.1937  & 51.7890  &  4.9086 &  59.3107 & {0.4814}  \\
        ~ & ISP+OSEDiff-s1 & {23.8067}  & 0.6872  & \cellcolor{yellow!40}{0.2988}  & \cellcolor{yellow!40}{0.1768}  &  52.0761 & 4.2011  & \cellcolor{yellow!40}{65.5805}  & \cellcolor{orange!40}{0.6793}  \\
        ~ & Ours-s1 &  {25.1264} & \cellcolor{yellow!40}{0.7092}  &  \cellcolor{red!40}{0.2546} &  \cellcolor{red!40}{0.1589} & \cellcolor{orange!40}{36.8671}  &  \cellcolor{orange!40}{4.1286} & \cellcolor{orange!40}{65.8881}   &  \cellcolor{yellow!40}{0.6723} \\
        \bottomrule
    \end{tabular}
    \label{tab:experiments}
\end{table*}

\subsection{Experimental Settings}
\label{sec:4.1}
\textbf{Training and Testing Datasets.}
RDDM is trained on LSDIR \cite{li2023lsdir} and a 10K-image subset of FFHQ \cite{karras2019style}. The RVAE is trained on a broader collection including DIV2K \cite{timofte2017ntire}, Flickr2K \cite{agustsson2017ntire}, LSDIR \cite{li2023lsdir}, DIV8K \cite{gu2019div8k}, and FFHQ \cite{karras2019style}. RAW-domain training pairs are synthesized following the pipeline in Sec.~\ref{section:image_generalization}. For evaluation, we employ both real-world and synthetic benchmarks. Real-world RAW testing is conducted on DND \cite{plotz2017benchmarking}, RealCapture \cite{li2024dualdn}, and SIDD \cite{SIDD_2018_CVPR}, covering diverse consumer cameras and mobile devices. Additionally, synthetic RAW benchmarks are constructed by applying our degradation method to HR images from DIV2K-Val \cite{timofte2017ntire}, RealSR \cite{cai2019toward}, and DRealSR \cite{wei2020component}.

\noindent\textbf{Compared Methods.} We compare RDDM against the best-performing traditional one-stage method and two-stage ISP$\rightarrow$IR models methods, as shown in Table \ref{tab:experiments}. For one-stage methods, we re-train JDnDmSR \citep{xingend}, SwinIR \citep{liang2021swinir}, and MambaIRv2 \citep{guo2025mambairv2} on the same RAW domain training dataset as our baseline. For two-stage ISP$\rightarrow$IR methods, we choose PIPNet \citep{a2021beyond} as the DN and DM module for ISP and diffusion-based IR methods as our Real-IR baselines, including StableSR \citep{wang2024exploiting}, DiffBIR \citep{lin2024diffbir}, PASD \citep{yang2024pixel}, SeeSR \citep{wu2024seesr}, SUPIR \citep{yu2024scaling}, and OSEDiff \citep{wu2024one}. Additionally, we conduct a comparative analysis with GAN-based Real-IR methods, including BSRGAN \cite{zhang2021designing}, Real-ESRGAN \cite{wang2021real}, LDL \cite{liang2022details}, and FeMaSR \cite{chen2022real}. The comparisons of RDDM against GAN-based Real-IR may be found in Appendix ~\ref{sec:A2}

\noindent\textbf{Evaluation Metrics.}
For a thorough assessment of the different methods, we utilize a variety of full-reference and non-reference evaluation metrics to test each method's image fidelity and generation quality. PSNR and SSIM \citep{wang2004image} (calculated on 3 channels) measure image fidelity, whereas LPIPS \citep{zhang2018unreasonable} and DISTS \citep{ding2020image} measure perceptual qualities based on reference images. FID \citep{heusel2017gans} assesses the distributional distance between the GT and the restored images. NIQE \citep{mittal2012making}, MUSIQ \citep{ke2021musiq}, and CLIPIQA \citep{wang2023exploring} are non-reference image generation quality measurements.

\noindent\textbf{Implementation Details.}
We train RDDM with the AdamW optimizer at a learning rate of $5e-5$. The entire training process spans 150000 steps with a batch size of 16. The rank of LoRA in the RVAE Encoder and the diffusion network is set to 4. We employ DAPE as the sRGB domain text prompts extractor. We set $\lambda_1=1$, $\lambda_2=1$ and $\lambda_3=3$   for all experiments.

\subsection{Comparisons with State-of-the-Arts}
\label{sec:4.2}

\begin{figure*}[t]
    \begin{minipage}[t]{0.48\linewidth}
    \captionsetup{type=figure}
      \centering
      \includegraphics[width=\linewidth]{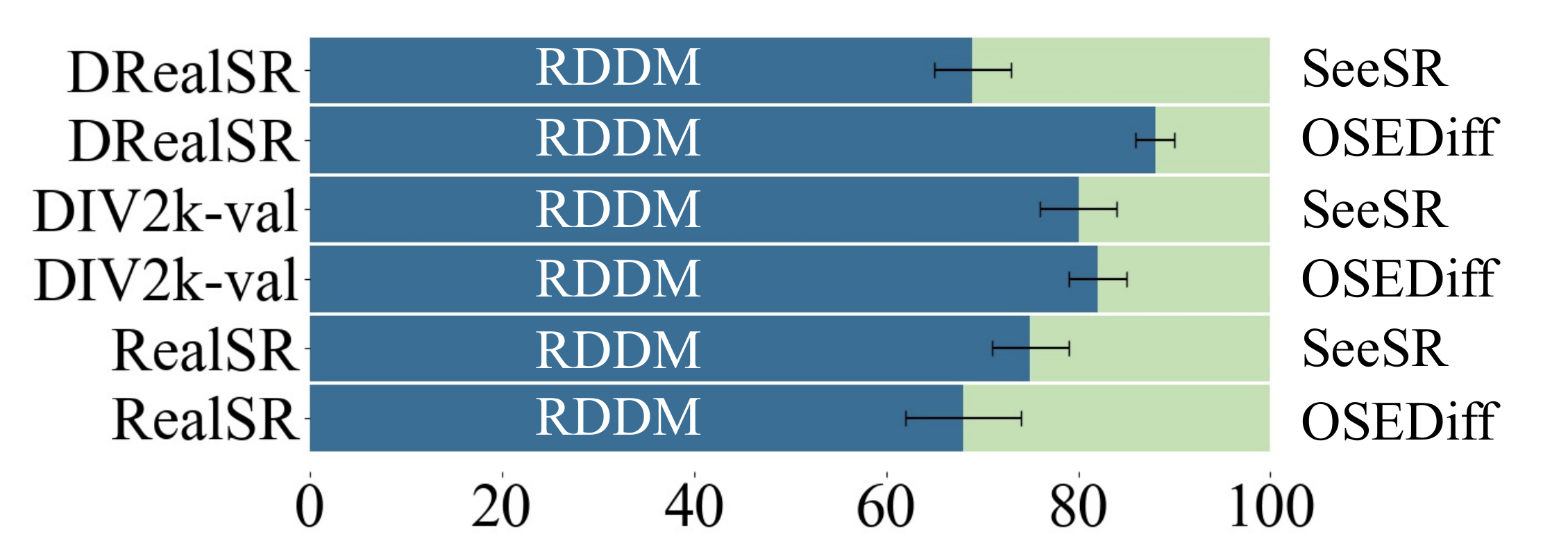}
      \caption{The user preference win rates of RDDM, compared to OSEDiff and SeeSR based on RealSR, DRealSR, and DIV2K-Val. We provide the 95\% confidence interval of the win rate based on five independent annotation rounds.}
      \label{fig:user_study}
    \end{minipage}
    \hfill
    \begin{minipage}[t]{0.48\linewidth}
    \captionsetup{type=table}
        \caption{Comparisons of Params and FLOPs between RDDM and its competing methods on input resolution of 512 $\times$ 512.}
        \centering
        \setlength{\tabcolsep}{3pt}
        \scriptsize
        \begin{tabular}{c|cc}
            \toprule
            \textbf{Method} & \textbf{Params(M)} & \textbf{FLOPs(G)} \\
            \midrule
             JDnDmSR & 78.2 & 54 \\
            SwinIR & 11.6 & 760 \\
            MambaIRv2 & 31.4 & 8260 \\
            ISP+StableSR-s200 & 1413 &  830 \\
            ISP+DiffBIR-s50 & 1673 &  1670 \\
            ISP+PASD-s20 & 1432 & 1590 \\
            ISP+SeeSR-s50 & 1622 & 1230 \\
            ISP+SUPIR-s50 & 4805 & 4100 \\
            ISP+OSEDiff-s1 & 1298 & 250 \\
            Ours-s1 & 1294 & 250 \\ 
            \bottomrule
        \end{tabular}
        \label{tab:params_flops}
    \end{minipage}
    \vspace{-10pt}
\end{figure*}

\textbf{Quantitative Comparisons on synthetic RAW benchmark.} 
Table \ref{tab:experiments} presents the quantitative comparisons on three synthetic test datasets. RDDM ranks in the top 3 for all metrics, including PSNR, SSIM, LPIPS, DISTS, NIQE, MUSIQ, CLIPIQA, and FID, across DIV2K-Val, DRealSR, and RealSR, except for PSNR on RealSR. JDnDmSR and SwinIR achieve slightly higher PSNR and SSIM on RealSR but significantly underperform on other metrics, particularly NIQE, MUSIQ, CLIPIQA, and FID, indicating weaker generative capabilities. RDDM matches diffusion-based methods in generative performance while outperforming them in image fidelity metrics like PSNR and SSIM. Moreover, RDDM maintains the highest efficiency among diffusion-based competitors in terms of Params and FLOPs as shown in Table~\ref{tab:params_flops}.

\begin{figure}[t]
  \centering
  \includegraphics[width=\textwidth]{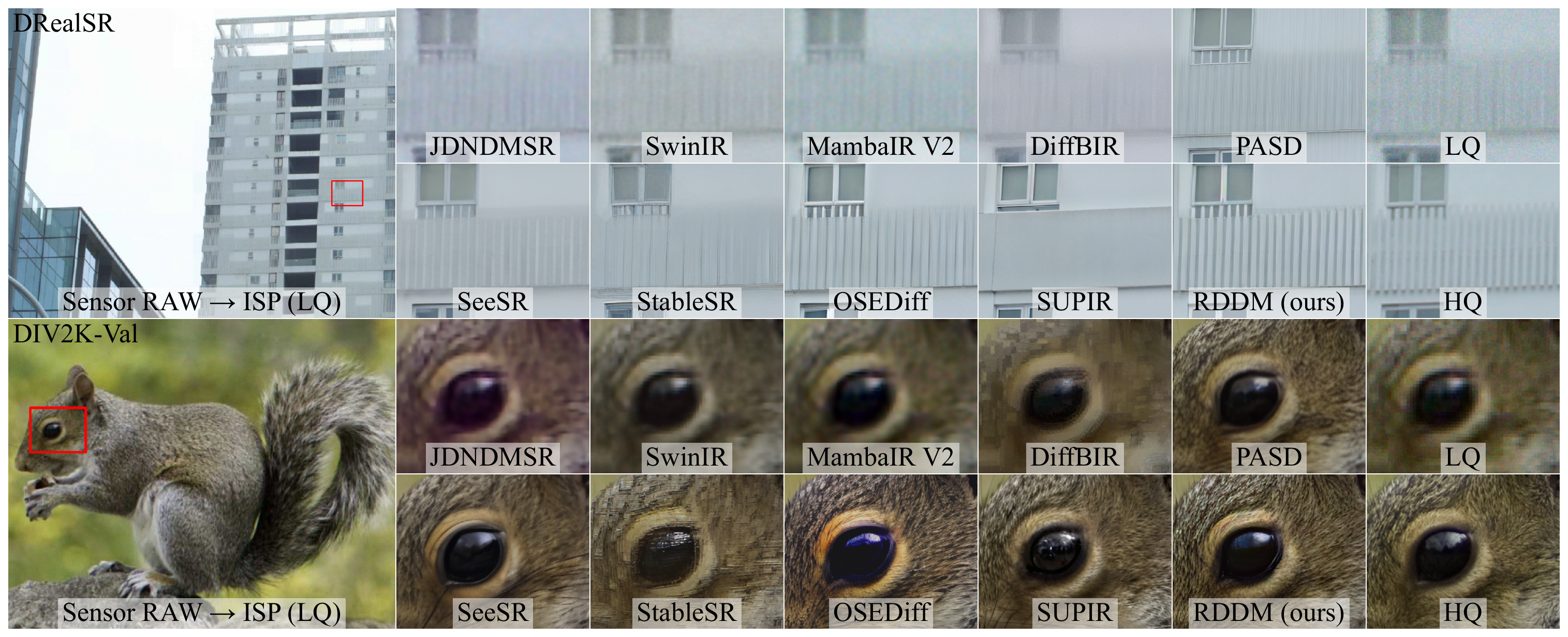}
  \caption{Qualitative comparison between RDDM and RAW domain one-stage method and two-stage ISP$\rightarrow$IR methods on DRealSR dataset and DIV2K-Val dataset.}
  \label{fig:experiment}
\end{figure}

\noindent\textbf{Qualitative Comparison on synthetic RAW benchmark.} Visual comparisons on DIV2K-Val and DRealSR demonstrate RDDM's superiority over existing one-stage and two-stage RAW-domain methods, as shown in Fig.\ref{fig:experiment}. While competitors like JDnDmSR and SwinIR suffer from blurriness and color bias, and diffusion-based models (e.g., SeeSR, OSEDiff) often fail to recover fine structures, RDDM reconstructs realistic, high-clarity textures. This confirms that RDDM effectively leverages RAW sensor signals to mitigate typical diffusion artifacts. More visualization comparison results are in the Appendix~\ref{sec:A3}. To further investigate the user preferences about these results, we conduct a user study on RealSR, DRealSR, and DIV2K-val test datasets, with 5 participants involved. For each set of comparison images, users select their preferred result. As shown in Fig. \ref{fig:user_study}, the results demonstrate that our method significantly outperforms state-of-the-art methods in terms of perceptual quality. 

\begin{figure*}[t]
  \centering
  \includegraphics[width=\textwidth]{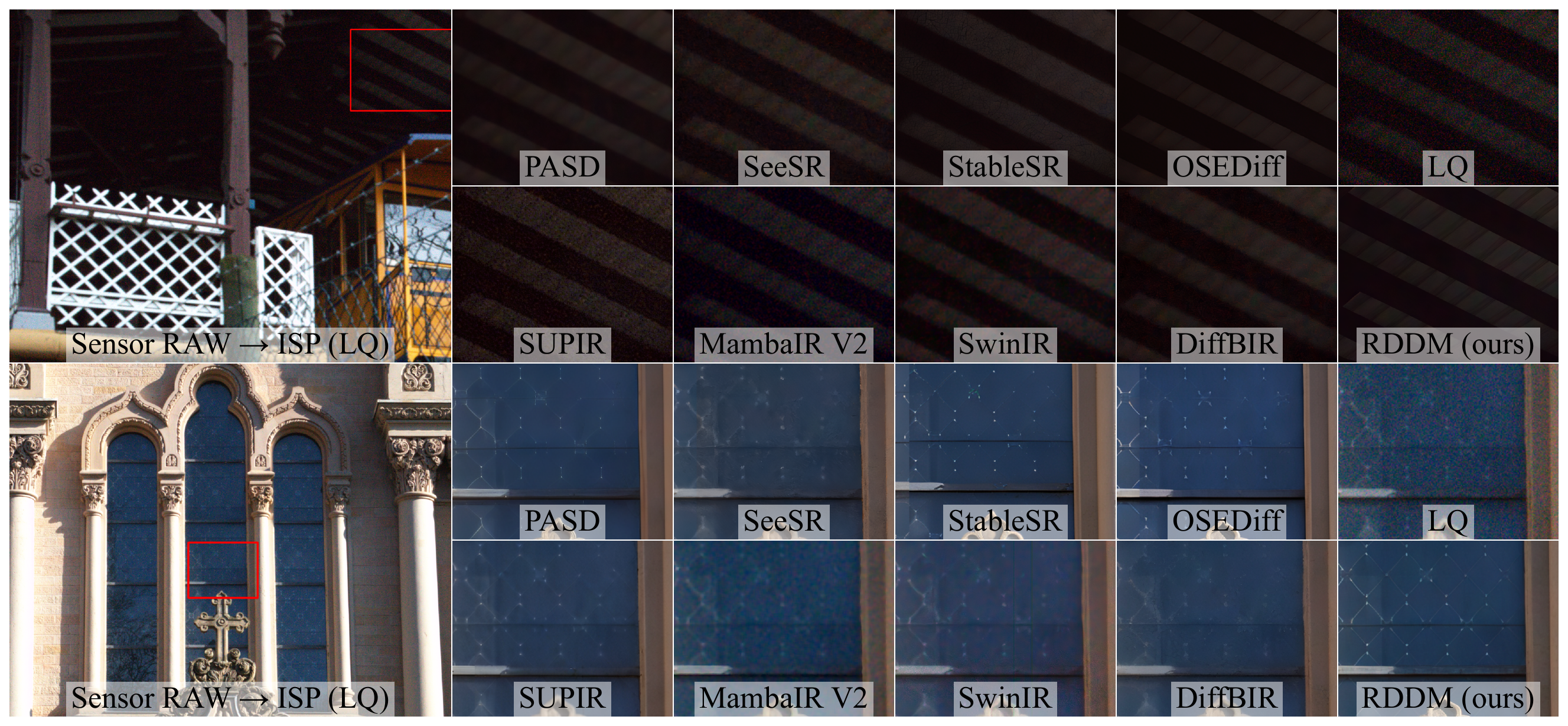}
  \caption{Qualitative comparison between RDDM and RAW domain one-stage method and two-stage ISP$\rightarrow$IR methods on the real RAW DND dataset.}
  \label{fig:experiment_DND}
\end{figure*}

\begin{figure*}[t]
  \centering
  \includegraphics[width=\textwidth]{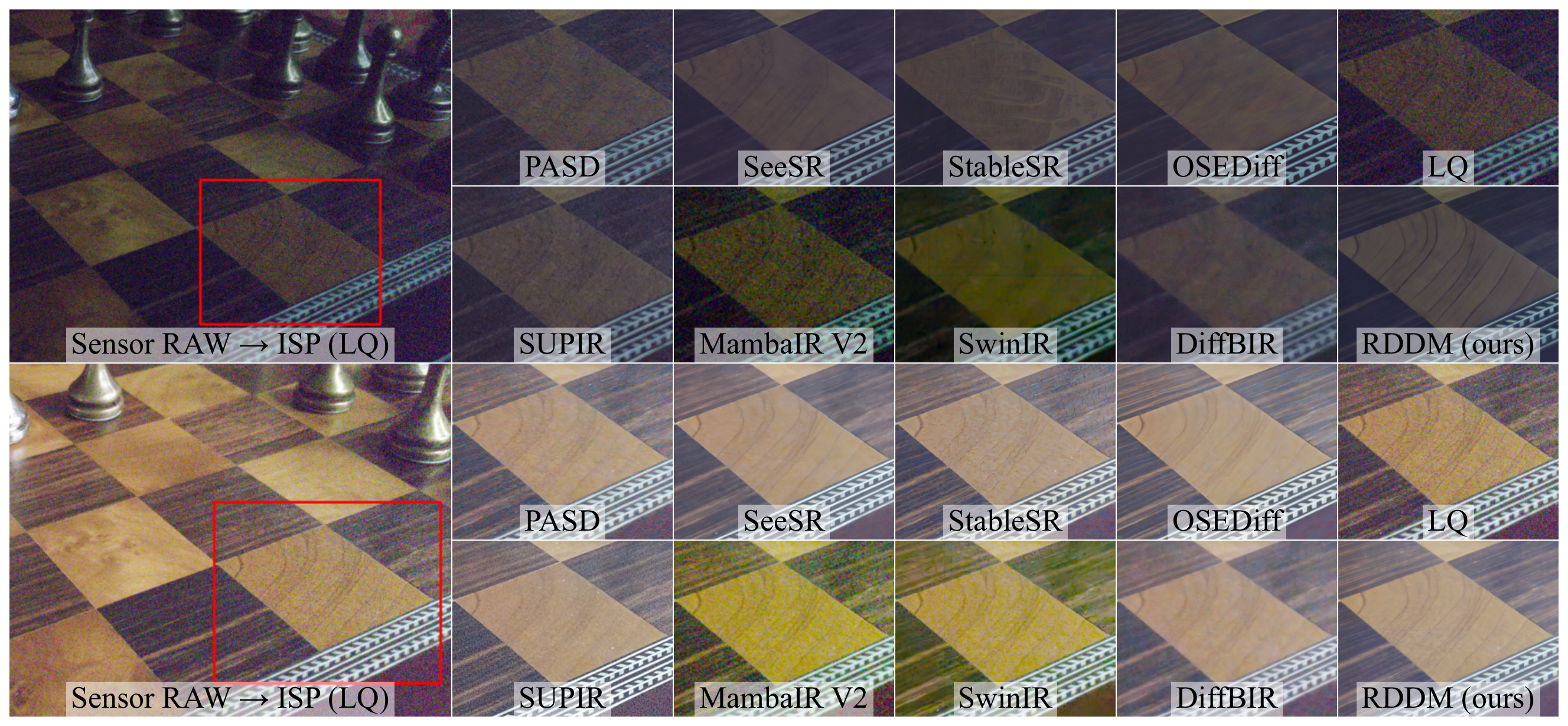}
  \caption{Qualitative comparison between RDDM and RAW domain one-stage method and two-stage ISP$\rightarrow$IR methods on the real RAW SIDD dataset.}
  \label{fig:experiment_SIDD}
\end{figure*}

\begin{figure*}[t]
  \centering
  \includegraphics[width=\textwidth]{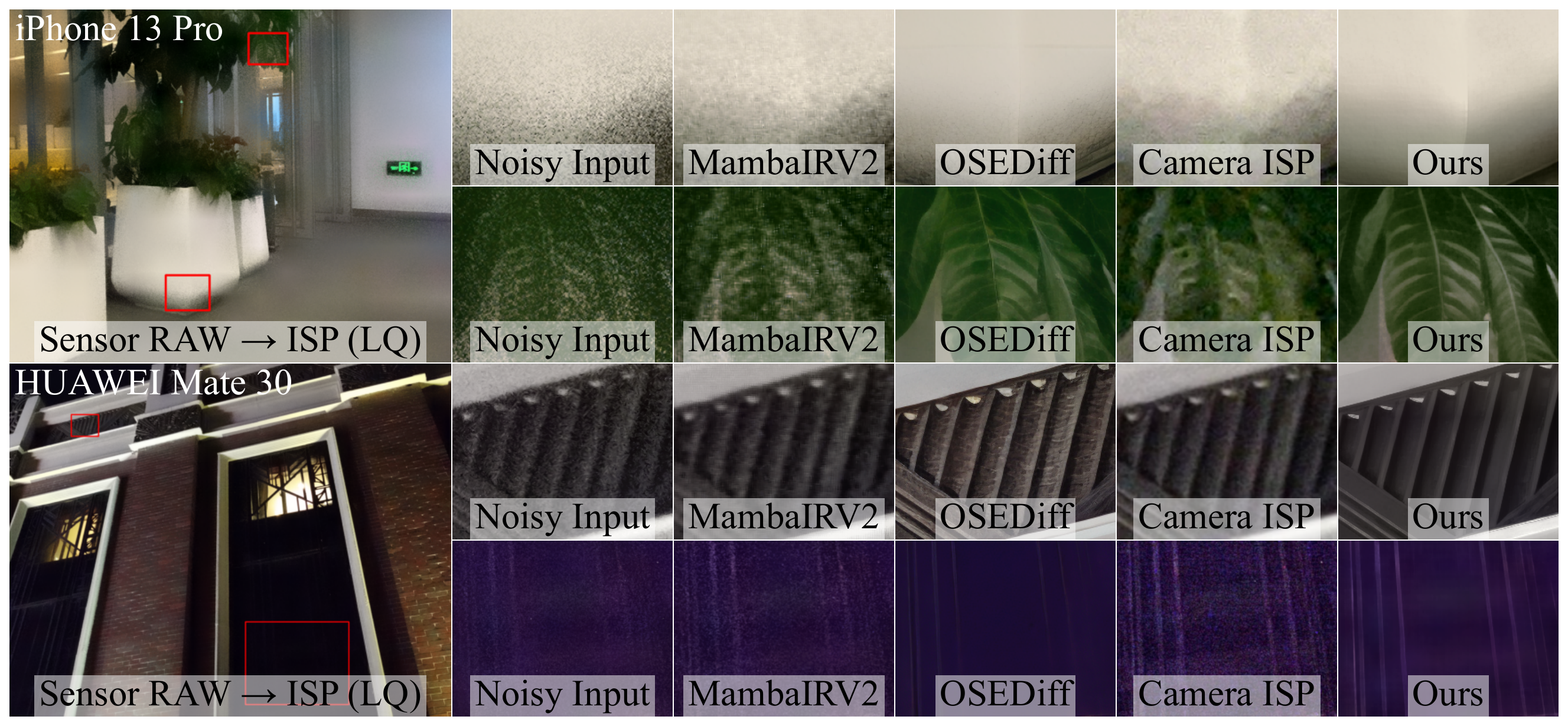}
  \caption{Qualitative comparison between RDDM and RAW domain one-stage method and two-stage ISP$\rightarrow$IR methods on the real RAW RealCapture dataset. The results in the last row are brightened to provide a clearer visualization of the restored details.}
  \label{fig:experiment_realcapture}
\end{figure*}

\noindent\textbf{Qualitative Comparison on real RAW benchmark.} Visual comparisons on the DND dataset (Fig.\ref{fig:experiment_DND}) highlight RDDM's ability to recover clear and realistic structures on complex surfaces, whereas competing methods often over-smooth textures or fail to suppress noise. As depicted in Fig.\ref{fig:experiment_SIDD}, on the multi-exposure SIDD dataset, RDDM consistently restores clean checkerboard patterns across varying exposure levels, avoiding the color deviations prevalent in one-stage models like MambaIR V2. Furthermore, RDDM exhibits robust generalization to unseen sensor patterns. When evaluated on the RealCapture dataset (RGBG Bayer pattern), RDDM successfully uncovers fine foliage and wall details, as illustrated in Fig.\ref{fig:experiment_realcapture}. In contrast, alternative methods suffer from severe residual noise, grid artifacts, and detail loss, confirming RDDM's superior cross-sensor adaptability.

\subsection{Ablation Study}

\begin{figure*}[t]
  \centering
  \includegraphics[width=\textwidth]{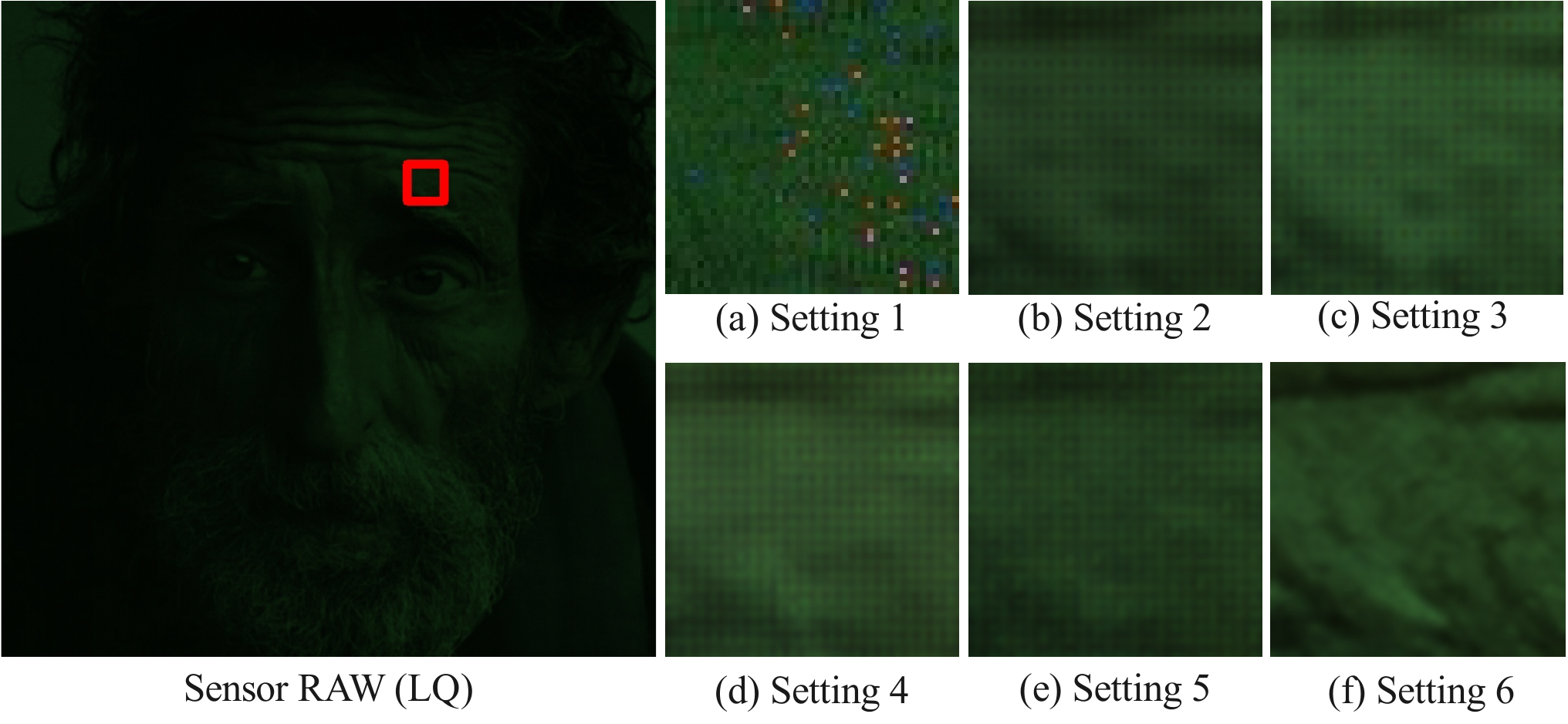}
  \caption{Qualitative comparison of different VAE settings on the RealSR benchmark. Setting 6 (ours) achieves the optimal performance.}
  \label{fig:ablation_vae_setting}
  \vspace{-5pt}
\end{figure*}

\begin{figure*}[t]
    \begin{minipage}[t]{0.48\linewidth}
    \captionsetup{type=figure}
      \centering
      \includegraphics[width=\textwidth]{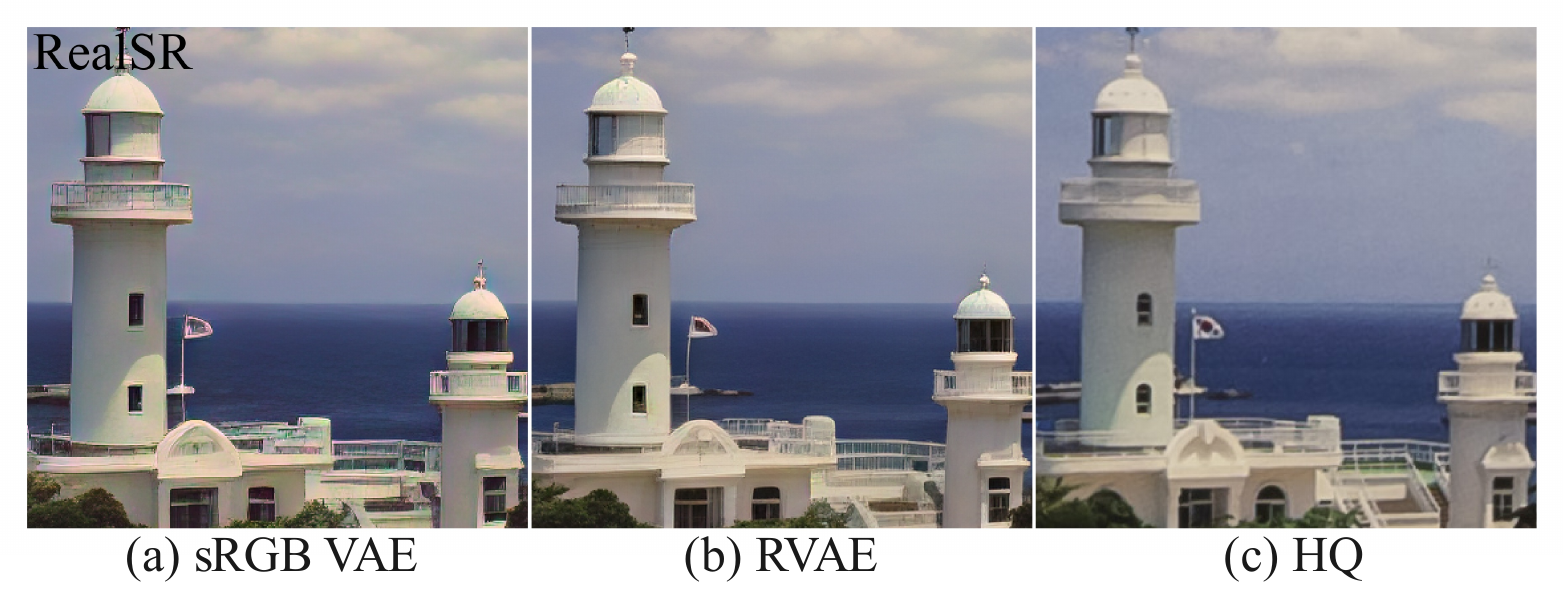}
      \caption{Qualitative comparison of sRGB VAE and RVAE on the RealSR benchmark.}
      \label{fig:ablation_vae}
    \end{minipage}
    \hfill
    \begin{minipage}[t]{0.48\linewidth}
    \captionsetup{type=table}
        \caption{Quantitative reconstruction performance of different VAE settings on RealSR.}
        \centering
            \setlength{\tabcolsep}{1.7mm}
            \scriptsize
        \begin{tabular}{@{}c|cccccccc@{}}
            \toprule
            \textbf{Setting} & \textbf{PSNR$\uparrow$} & \textbf{SSIM$\uparrow$} & \textbf{LPIPS$\downarrow$} & \textbf{FID$\downarrow$} \\ \midrule
            1         & 25.2625         & 0.8017        & 0.1719                        & 41.8118       \\
            2           & 27.1487         & 0.7035        & 0.3068                    & 48.4675         \\
            3         &  27.6892 & 0.7176  &  0.2787  & 38.2007  \\ 
            
            4         & 26.3147     &  0.6703       &   0.3967      &  53.4004       \\
            5         & 25.4509      &  0.6277       &  0.4533      &  64.2444       \\
            6         & 32.5424          & 0.9082        & 0.0533                      & 11.6156        \\
            \bottomrule
        \end{tabular}
        \label{tab:ablation_modules_RVAE}
    \end{minipage}
    \vspace{-10pt}
\end{figure*}

\textbf{The Importance of RVAE.} To identify the optimal RVAE architecture and training protocol, we conduct a comprehensive evaluation of six adaptation strategies for RAW-domain reconstruction. The qualitative and quantitative results are summarized as follows:

(i) \textit{Baseline Adaptations (Settings 1–2):} directly employing a pre-trained sRGB VAE or training from scratch yields either severe degradation or persistent mosaic artifacts (Fig.\ref{fig:ablation_vae_setting}a-b).

(ii) \textit{LoRA-based Fine-tuning (Settings 3–5):} furthermore, fine-tuning only the encoder, decoder, or both via LoRA proves insufficient to bridge the domain gap between sRGB and RAW, as mosaic patterns remain prominent (Fig.\ref{fig:ablation_vae_setting}c-e).

(iii) \textit{Proposed Joint Training (Setting 6):} the optimal strategy involves freezing a pre-trained linear domain decoder while jointly optimizing the encoder and diffusion network via LoRA. This configuration effectively eliminates artifacts, as demonstrated in Fig.\ref{fig:ablation_vae_setting}(f).

Quantitative metrics in Table \ref{tab:ablation_modules_RVAE} further validate that Setting 6 achieves the highest reconstruction fidelity. Furthermore, End-to-end evaluations within the RDDM framework (Table~\ref{tab:ablation_modules_RVAE_RDDM}) confirm that our RVAE strategy provides the most robust foundation for high-quality RAW-to-linear restoration, significantly outperforming alternative adaptation protocols. Moreover, Fig.\ref{fig:ablation_vae} confirms that RVAE effectively mitigates chromatic deviations.

\begin{table}[t]
\captionsetup{type=table}
\caption{Quantitative performance of RDDM with different VAE settings on the RealSR dataset.}
\centering
    \scriptsize
\begin{tabular}{@{}c|cccccccc@{}}
    \toprule
    \textbf{Setting} & \textbf{PSNR$\uparrow$} & \textbf{SSIM$\uparrow$} & \textbf{LPIPS$\downarrow$} & \textbf{DISTS$\downarrow$} &  \textbf{FID$\downarrow$} & \textbf{NIQE$\downarrow$} & \textbf{MUSIQ$\uparrow$} & \textbf{CLIPIQA$\uparrow$} \\ \midrule
    1         &   20.5609   &   0.5797   &   0.3930    &   0.2071   &   73.5659   &  4.0762    &    62.7875   &   0.6431     \\
    2         & 22.3752  & 0.6484  & 0.4785  & 0.3307  & 138.4347 & 6.5993 & 47.2304 &   0.4233   \\
    3         & 21.0398  & 0.6115  & 0.5710  &  0.4235 & 143.3710 & 10.5422   &  37.3569 &   0.3745\\ 
    4         & 21.3850    &  0.6080    &  0.4250     &  0.2901    &   108.3618   & 7.5271     &   52.8636    &   0.2931    \\
    5         &   21.3886   &   0.5732   &   0.4911    &   0.3166   &   135.6228   &   7.5537   &    38.7119   & 0.1483 \\
    6         &  25.1264    &  0.7092    &   0.2546    &   0.1589   &    36.8671  &   4.1286   &   65.8881    &    0.6723    \\
    \bottomrule
\end{tabular}
\label{tab:ablation_modules_RVAE_RDDM}
\end{table}

\begin{table*}[t]
    \caption{Comparison of CMB LoRA and All-in-One LoRA on RealSR benchmark.}
    \vspace{3mm}
    \centering
        \setlength{\tabcolsep}{1.7mm}
        \scriptsize
    \begin{tabular}{@{}c|cccccccc@{}}
        \toprule
        \textbf{} & \textbf{PSNR$\uparrow$} & \textbf{SSIM$\uparrow$} & \textbf{LPIPS$\downarrow$} & \textbf{DISTS$\downarrow$} & \textbf{FID$\downarrow$} & \textbf{NIQE$\downarrow$} & \textbf{MUSIQ$\uparrow$} & \textbf{CLIPIQA$\uparrow$}  \\ \midrule
        All-in-One LoRA  &  24.5355     &     0.7003     &    0.2550     &    0.1555       &        40.6767       &     4.1165     &     65.0143       &     0.6660    \\
        CMB-LORA          &  25.1264         & 0.7092        & 0.2546         & 0.1589              & 36.8671         & 4.1286           & 65.8881        & 0.6723        \\
        \bottomrule
    \end{tabular}
    \label{tab:ablation_modules_CMB}
\end{table*}

\textbf{The Necessity of CMB-LoRA.} We investigate the parameter-sharing strategy within the CMB-LoRA module by comparing our pattern-specific design against a shared-weight baseline. As reported in Table~\ref{tab:ablation_modules_CMB}, the shared configuration leads to performance degradation across diverse sensor layouts. This decline stems from the inherent spatial disparities among heterogeneous Bayer patterns (e.g., RGGB, BGGR, and RGBG). While a single weight matrix struggles to capture such diverse local structural characteristics, our pattern-specific design provides the necessary flexibility to model the unique statistical distributions of each RAW format, ensuring robust adaptation across heterogeneous sensors.

 \begin{figure*}[t]
  \centering
  \includegraphics[width=\textwidth]{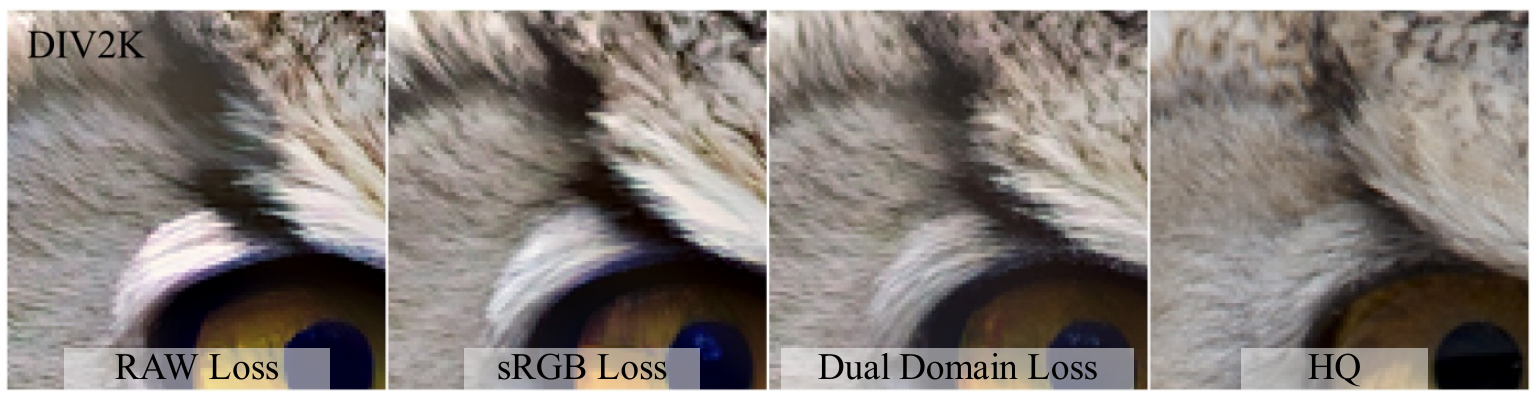}
  \caption{Qualitative comparison of RAW, sRGB, and dual domain loss on the DIV2K-Val.}
  \label{fig:ablation_losses}
\end{figure*}

\begin{table*}[t]
    \caption{Comparison of different domain losses on the DIV2K benchmark.}
    \centering
        \setlength{\tabcolsep}{1.7mm}
        \scriptsize
    \begin{tabular}{@{}cc|cccccccc@{}}
        \toprule
        \textbf{RAW Loss} & \textbf{sRGB Loss} & \textbf{PSNR$\uparrow$} & \textbf{SSIM$\uparrow$} & \textbf{LPIPS$\downarrow$} & \textbf{DISTS$\downarrow$} & \textbf{FID$\downarrow$} & \textbf{NIQE$\downarrow$} & \textbf{MUSIQ$\uparrow$} & \textbf{CLIPIQA$\uparrow$}  \\ \midrule
        \checkmark         & $\times$          & 22.9510          &  0.6096        & 0.2794         & 0.1337              & 36.3172         & 3.4637           & 66.1391        & 0.7298         \\
        $\times$         & \checkmark          &      23.3129     &    0.6237     &     0.2635     &        0.1216       &     29.5131     &      3.3352      &    64.7748     &     0.6637    \\
        
        \checkmark         & \checkmark        &  23.7416         & 0.6296        & 0.2540         & 0.1197    & 23.8028  & 3.3627     & 65.4202        & 0.6737        \\
        \bottomrule
    \end{tabular}
    \label{tab:ablation_modules_losses}
\end{table*}

\textbf{The Effectiveness of Dual Domain Loss.} We evaluate the impact of domain-specific losses in Table~\ref{tab:ablation_modules_losses}. Exclusively RAW-domain training (Setting 1) yields high non-reference scores (e.g., MUSIQ, CLIPIQA) but introduces chromatic artifacts due to the lack of color constraints. Conversely, sRGB-only optimization (Setting 2) ensures color consistency but sacrifices structural fidelity by discarding sensor-level signals. By imposing dual-domain supervision, RDDM achieves an optimal trade-off between structural authenticity and visual realism. As illustrated in Fig.\ref{fig:ablation_losses}, while this strategy shows a marginal decline in reference-less metrics, which often over-reward texture intensity, it significantly improves global color stability and alignment with the ground truth, prioritizing holistic fidelity over local high-frequency noise.

\begin{table*}[t]
    \caption{Comparison of different text prompt extractors on the RealSR benchmark.}
    \centering
        \setlength{\tabcolsep}{1.7mm}
        \scriptsize
    \begin{tabular}{@{}c|cccccccc@{}}
        \toprule
        \textbf{} & \textbf{PSNR$\uparrow$} & \textbf{SSIM$\uparrow$} & \textbf{LPIPS$\downarrow$} & \textbf{DISTS$\downarrow$} & \textbf{FID$\downarrow$} & \textbf{NIQE$\downarrow$} & \textbf{MUSIQ$\uparrow$} & \textbf{CLIPIQA$\uparrow$}  \\ \midrule
        sRGBPE         &    24.7811       &     0.7201 
    &     0.2495       &       0.1601         &     38.1092      &     4.2538        &     65.4506      & 0.6579     \\
        ISPPE          &      24.7816      &    0.7201      & 0.2495      &        0.1600       &    38.0969      &     4.2539        &     65.4526     &    0.6579      \\
        RPE          &  25.1264 & 0.7092  &  0.2546 &  0.1589 & 36.8671  & 4.1286 & 65.8881 &  0.6723        \\
        \bottomrule
    \end{tabular}
    \label{tab:ablation_modules_RPE}
\end{table*}

\textbf{The Comparison of Text Prompt Extractors.} To evaluate the efficacy of our proposed RPE, we compare it against two alternative prompt-extraction strategies: (1) Direct sRGB-Extractor, which utilizes a pre-trained sRGB model on RAW data; and (2) ISP-based Conversion, which extracts prompts from RAW images pre-processed by a standard ISP. As summarized in Table \ref{tab:ablation_modules_RPE}, the proposed RPE consistently outperforms both baselines across all metrics, including reconstruction fidelity (PSNR, DISTS) and perceptual quality (FID, NIQE, MUSIQ, CLIPIQA). By generating more precise prompts, RPE effectively activates the rich generative priors of the diffusion model, facilitating high-fidelity restoration and superior detail synthesis that standard or ISP-processed extractors fail to achieve.

\section{Conclusion}

In this paper, we present RDDM, a pioneering paradigm for Real-IR that operates directly from sensor RAW data. By integrating the robust generative priors of the diffusion model, RVAE with a specialized CMB-LoRA module, RDDM effectively addresses the inherent limitations of traditional sRGB-based restoration. Unlike conventional methods that struggle with the information loss incurred by irreversible ISP pipelines, our approach leverages the unprocessed native signal of RAW images to recover superior structural details and textures. While the current 8$\times$ down-sampling architecture inherited from Stable Diffusion 2.1 imposes certain constraints on fine-scale fidelity. Future research will focus on developing VAE encode strategies to further preserve sub-pixel spatial information and push the boundaries of high-fidelity Real-IR.

\newpage

\bibliography{iclr2026_conference}
\bibliographystyle{iclr2026_conference}

\clearpage

\appendix
\section{Appendix}\label{sec:app}

The appendix provides additional technical details, experimental results, and visual comparisons to support the main paper. Specifically, it includes:
\begin{itemize}
    \item Additional demonstrations regarding ISP and inverse ISP (referring to Sections ~\ref{sec:3.1} and ~\ref{section:image_generalization} in the main paper).
    \item Comparisons against GAN-based methods (referring to Section ~\ref{sec:4.2} in the main paper).
    \item More qualitative comparisons (referring to Section ~\ref{sec:4.2} in the main paper).
    \item Visualizations of some examples of the synthetic RAW data. (referring to Section ~\ref{section:image_generalization} in the main paper).
\end{itemize}

\subsection{ISP and Inverse ISP} \label{sec:A1}

\begin{figure*}[t]
  \centering
  \includegraphics[width=\textwidth]{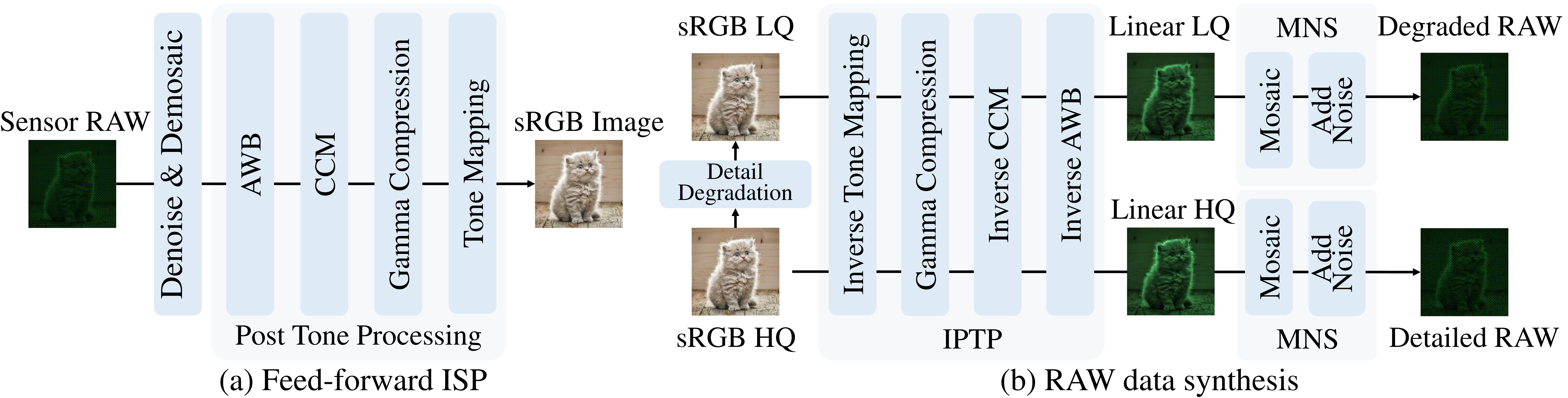}
  \caption{The ISP pipeline and the data synthesis pipeline.}
  \label{fig:ISP_data_synthesis_detail}
\end{figure*}

\textbf{ISP and Inverse ISP.} The RAW sensor data obtained by a camera is fundamentally different from the sRGB images, which closely resemble human visual perception. Indeed, it is necessary to process RAW images through an ISP to obtain the final sRGB images. Fig.\ref{fig:ISP_data_synthesis_detail} illustrates the processing pipelines of ISP and InverseISP. We modify the InverseISP of \cite{brooks2019unprocessing} and propose a differentiable ISP approach that can map the model's output to the sRGB domain for optimization. During the ISP process, denoising and demosaicing are ill-posed problems, and we optimize these two tasks jointly with the Real-IR task. Above all, in the ISP process, the Automatic White Balance (AWB) algorithm multiplies the red and blue channels by gains to produce an image that appears to be lit under “neutral” illumination, which can be formulated as Equation \ref{equation:AWB}:

\begin{equation}
\begin{aligned}
\label{equation:AWB}
    x_{L}^{awb} &= X_{L}^{nm} \odot M_{gain} \\
    x_{L}^{nm} &= X_{L}^{awb} \odot \frac{1}{M_{gain}}
\end{aligned}
\end{equation}
where $x_{L}^{awb}$, $X_{L}^{nm}$, and $M_{gain}$ are the linear-domain image processed by AWB, the linear image processed by denoising and demosaicing, and the pixel-wise channel gain, respectively.  

Secondly, the color correction algorithm converts its own “camera space” sRGB color measurements to sRGB values by a color correction matrix as shown in Equation \ref{equation:CCM}:

\begin{equation} 
\begin{aligned}
\label{equation:CCM}
    x_{L}^{ccm} &= x_{L}^{awb} \times M_{c} \\
    x_{L}^{awb} &= x_{L}^{ccm} \times M_{c}^{-1}
\end{aligned}   
\end{equation}
where $x_{L}^{ccm}$, $M_{c}$ are the linear-domain image processed by the CCM algorithm and the color correction matrix, respectively. 

\noindent Thirdly, since humans are more sensitive to gradations in the dark areas of images, gamma compression is typically used to allocate more bits of dynamic range to low-intensity pixels as stated in Equation \ref{gamma compression}:

\begin{equation} 
\label{gamma compression}
\begin{aligned}
    X_{L}^{gamma} &= max(x_{L}^{ccm}, \epsilon) ^{1/2.2} \\
    X_{L}^{ccm} &= max(x_{L}^{gamma}, \epsilon) ^{2.2}
\end{aligned}   
\end{equation}
Note that we set $\epsilon=10^{-8}$ to prevent numerical instability during training. 

\noindent InverseISP is the inverse process of ISP, where the mosaicing algorithm acquires the RAW image $X_{L}^{raw} \in \mathcal{R}^{h \times w \times 1}$ with a CFA by extracting the corresponding pixel values from the three channels. The noise in RAW images mainly comes from two sources: photon arrival statistics (shot noise) and imprecision in the readout circuitry (read noise). We can approximate these two types of noise as a single heteroscedastic Gaussian distribution defined in Equation \ref{read and shot noise}:

\begin{equation} \label{read and shot noise}
    y \sim  \mathcal{N}(\mu = 0, \sigma^2 = \lambda_{shot}X_{L}^{mosaic} + \lambda_{read}) 
\end{equation}
The linear-domain image is obtained by Equation \ref{RAW noise}:
\begin{equation} \label{RAW noise}
    X_{L}^{RAW} = X_{L}^{mosaic} + y
\end{equation}
where $X_{L}^{mosaic}$ is the RAW image processed by mosaic algorithm and $y$ is the noise intensity added onto $X_{L}^{mosaic}$ to obtain the Sensor RAW $X_{L}^{RAW}$. $\lambda_{shot}$ and $\lambda_{read}$ are the function of ISO light sensitivity level.

\subsection{Quantitative Comparisons with GAN-based Methods}
\label{sec:A2}

\begin{table*}[t]
    \centering
    \setlength{\tabcolsep}{3pt}
    \scriptsize
    \begin{tabular}{c|c|ccccccccc}
        \toprule
        \textbf{Dataset} & \textbf{Method} & \textbf{PSNR$\uparrow$} & \textbf{SSIM$\uparrow$} & \textbf{LPIPS$\downarrow$} & \textbf{DISTS$\downarrow$} & \textbf{FID$\downarrow$} & \textbf{NIQE$\downarrow$} & \textbf{MUSIQ $\uparrow$} & \textbf{CLIPIQA $\uparrow$}\\
        \midrule
        \multirow{5}*{DIV2K-Val} & ISP+BSRGAN & \textbf{24.4124}  & \textbf{0.6500}  & 0.4174  & 0.2173  & 33.8937  & 5.1222  & 44.2923  & 0.4580 \\
        ~ & ISP+Real-ESRGAN & 24.2373  &  0.6426 &  0.3888 & 0.2007  & 35.3941 &  4.8658 &  48.6062  &  0.5157 \\
        ~ & ISP+LDL &  24.2922 &  0.6243 & 0.4684  & 0.2540  &  34.9855 & 5.5153  & 29.8086  & 0.3714  \\
        ~ & ISP+FeMaSR & 23.5889  & 0.6066  & 0.4347  & 0.2273  & 34.9390  & 5.7403  & 43.3786  & 0.5187  \\
        ~ & Ours & {23.7416} & {0.6296} &  \textbf{0.2540} & \textbf{0.1197}  &  \textbf{23.8028}  & \textbf{3.3627}  & \textbf{65.4202}    & \textbf{0.6737}  \\
        \midrule
        \multirow{5}*{DRealSR} & ISP+BSRGAN & \textbf{30.1134}  & \textbf{0.8282}  & 0.3196  &  0.1982 & 20.5962  &  5.6748 & 42.4667  & 0.5105 \\
        ~ & ISP+Real-ESRGAN &  29.1309 & 0.8055  & 0.3361  &  0.2042 & 23.5094 & 5.5440  & 46.0976  &  0.5548 \\
        ~ & ISP+LDL & 29.9312  &  0.7967 &  0.3640 &  0.2173 & 20.9155  & 5.9973  &  29.3485 &  0.4047 \\
        ~ & ISP+FeMaSR &  28.1176 & 0.7460  &  0.4007 & 0.2356  & 24.3254  & 6.3530  & 41.3755 &  0.5844 \\
        ~ & Ours &  {28.3495} &  {0.7892} & \textbf{0.2719}  & \textbf{0.1649}  &  \textbf{17.4825} & \textbf{4.6852}  & \textbf{57.0696}  &  \textbf{0.7035} \\
        \midrule
        \multirow{5}*{RealSR} & ISP+BSRGAN & \textbf{27.0999}  & \textbf{0.7648}  & 0.3190  & 0.1984  & 50.7266  & 5.0958  & 50.2340   & 0.5052 \\
        ~ & ISP+Real-ESRGAN & 26.3844  & 0.7474  & 0.3332  &  0.1995 & 49.3939 &  4.5754 &  55.7670  &  0.5498 \\
        ~ & ISP+LDL & 26.5426  & 0.7186  & 0.3871  & 0.2141  & 53.1533  & 5.2573  & 33.9660    & 0.3656  \\
        ~ & ISP+FeMaSR & 25.5820 & 0.6898 & 0.3891 & 0.2278  &  53.7582 & 5.8086  & 51.1097 &  0.5917 \\
        ~ & Ours &  {25.1264} & {0.7092}  &  \textbf{0.2546} &  \textbf{0.1589} & \textbf{36.8671}  &  \textbf{4.1286} & \textbf{65.8881}   &  \textbf{0.6723} \\
        \bottomrule
    \end{tabular}
    \caption{Quantitative comparison with different methods on both synthetic benchmarks. $\downarrow$ presents the smaller the better, $\uparrow$ presents the bigger the better. Our method RDDM exceeds its competing models in terms of image fidelity, while maintaining a satisfactory image generation ability approximately equivalent to the other baselines.}
    \label{tab:experiments_GAN}
\end{table*}

\begin{figure}[t]
  \centering
  \includegraphics[width=\linewidth]{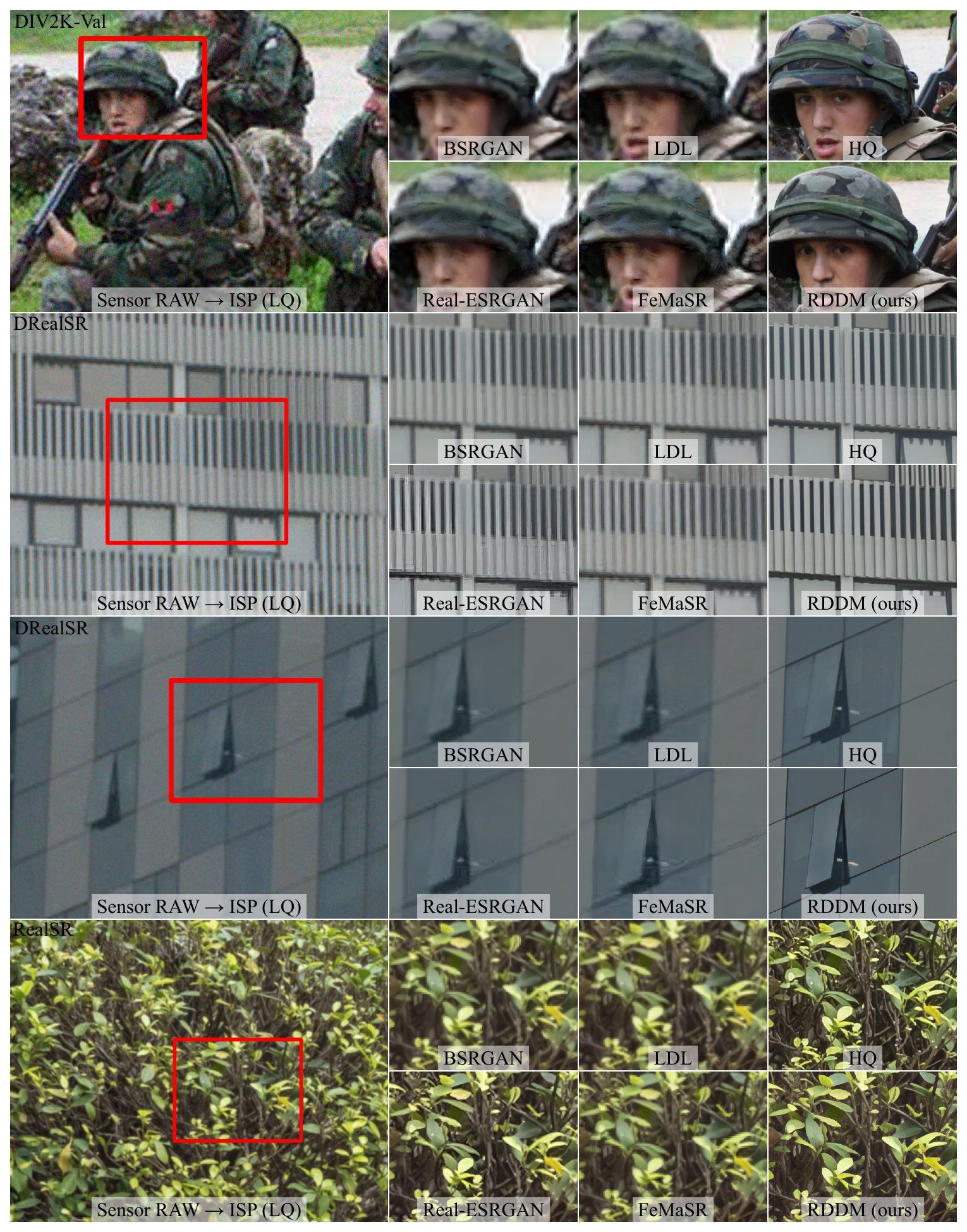}
  \caption{Qualitative comparison between RDDM and GAN-based methods on DIV2K-Val, DRealSR and RealSR dataset.}
  \label{fig:experiment_gan_visual}
\end{figure}

\noindent We compare RDDM against four two-stage ISP$\rightarrow$GAN-based-IR methods. For the first ISP stage, we again use PIPNet \cite{a2021beyond} as the DN and DM module; for the second IR stage, we use BSRGAN \cite{zhang2021designing}, Real-ESRGAN \cite{wang2021real}, LDL \cite{liang2022details}, and FeMaSR \cite{chen2022real} as our baselines. The results are shown in Table \ref{tab:experiments_GAN}. Admittedly, two-stage methods involving GAN-based IR models demonstrate better fidelity results as expected, i.e., higher PSNR and SSIM metrics. However, their image generation capability is far behind RDDM. Visualization comparisons are shown in Fig.~\ref{fig:experiment_gan_visual}.

\subsection{More Qualitative Comparisons}
\label{sec:A3}

\begin{figure}[t]
  \centering
  \includegraphics[width=\linewidth]{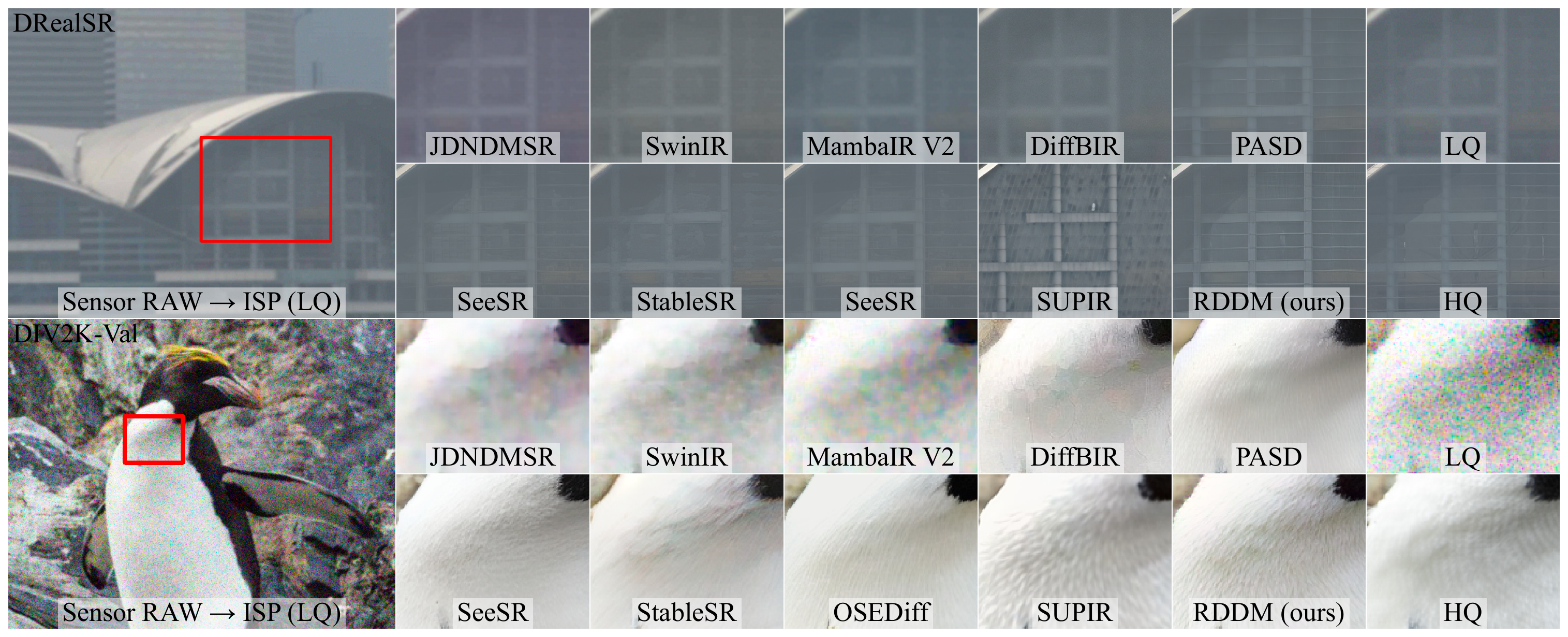}
  \caption{Qualitative comparison between RDDM and RAW domain one-stage method and two-stage ISP$\rightarrow$IR methods on DRealSR dataset and DIV2K-Val dataset.}
  \label{fig:experiment_appendix}
\end{figure}

\noindent To further demonstrate the perceptual superiority of our approach, we provide extended qualitative comparisons in Fig.\ref{fig:experiment_appendix}. These comparisons encompass both RAW-domain one-stage methods and representative two-stage pipelines (i.e., ISP followed by sRGB domain IR). The visual results across diverse challenging scenarios consistently showcase that RDDM achieves superior texture restoration and color fidelity, effectively mitigating the artifacts typically found in traditional multi-stage processing.

\subsection{Examples of Synthetic RAW Data}
\label{sec:A7}

\begin{figure}[H]
  \centering
  \includegraphics[width=\textwidth]{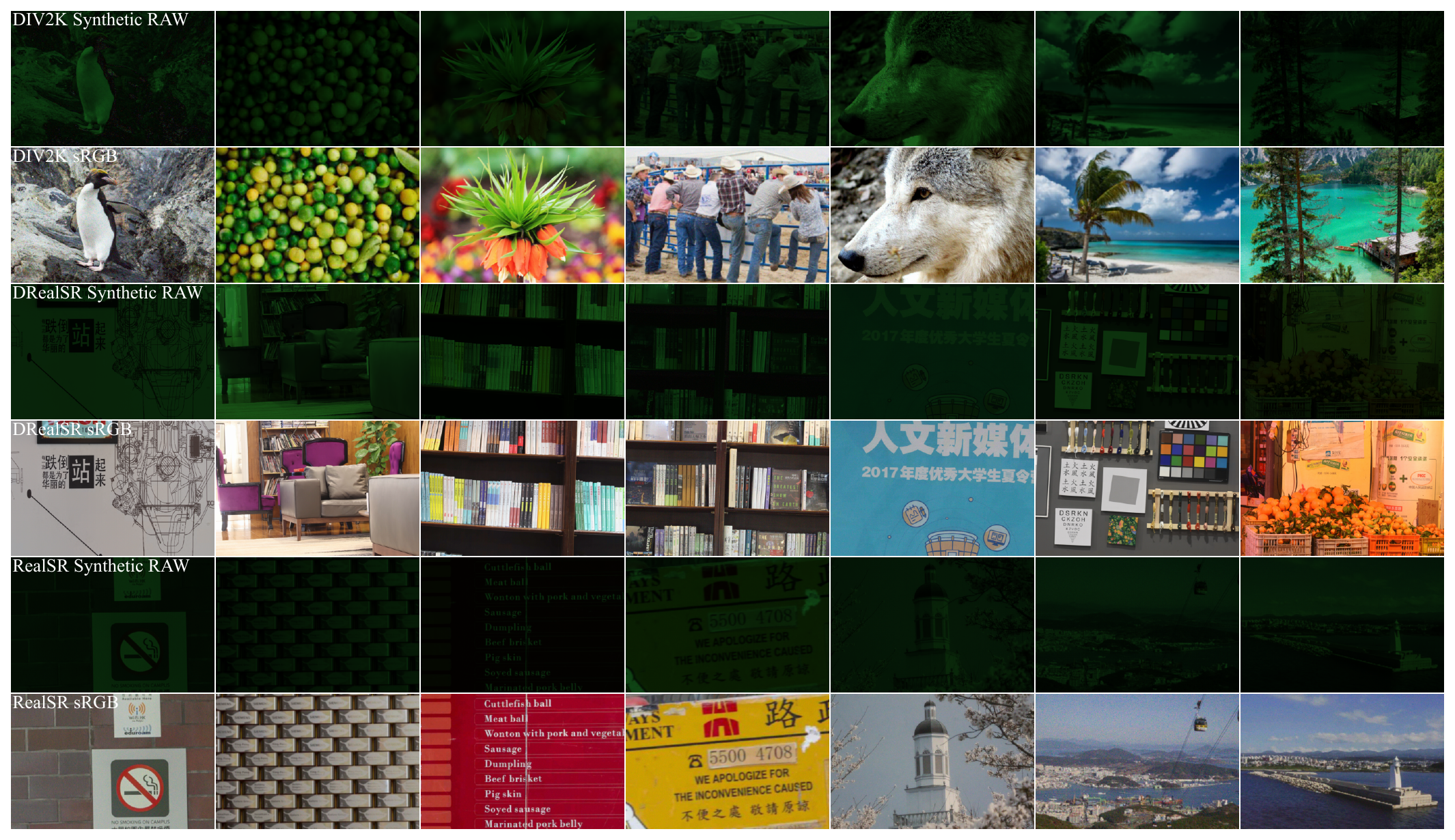}
  \caption{Examples of synthetic RAW data produced by our data synthesis pipeline. From top to bottom, each row represents the RAW data from DIV2K, the corresponding sRGB data from DIV2K, the RAW data from DRealSR, the corresponding sRGB data from DRealSR, the RAW data from RealSR and the corresponding sRGB data from RealSR.}
  \label{fig:synthetic_raw_imgs_examples}
\end{figure}

\noindent Following the data synthesis pipeline detailed in Fig.\ref{fig:ISP_data_synthesis_detail}, we provide representative examples of our synthesized RAW images in Fig.\ref{fig:synthetic_raw_imgs_examples}. These samples visually demonstrate that our pipeline effectively simulates realistic sensor noise and Bayer mosaics. By bridging the gap between synthetic degradations and real-world RAW characteristics, these generated images provide high-quality supervision, which is fundamental to the robust performance of RDDM.

\end{document}